\documentclass[%
reprint,superscriptaddress,amssymb,amsmath,aps,showpacs,10pt,longbibliography,pra,floatfix%
]{revtex4-2}
\usepackage{graphicx}
\pdfoutput=1

\graphicspath{{figures/}{../figures/}}
\usepackage{color}
\usepackage[colorlinks,bookmarks=false,citecolor=darkblue,linkcolor=red,urlcolor=blue]{hyperref}
\def\cred{
}
\definecolor{darkred}{rgb}{0.7,0.0,0.0}
\definecolor{darkblue}{rgb}{0,0.02,0.45}

\newcommand{\Dv}{\mathbf D}

\begin{document}


\title{Toward cubic symmetry for Ir$^{4+}$: structure and magnetism of the antifluorite K$_2$IrBr$_6$}

\author{Nazir Khan}
\email{nazirkhan91@gmail.com}
\affiliation{Experimental Physics VI, Center for Electronic Correlations and Magnetism, Institute of Physics, University of Augsburg, 86135 Augsburg, Germany}

\author{Danil Prishchenko}
\affiliation{Ural Federal University, Mira Str. 19, 620002 Ekaterinburg, Russia}

\author{Mary H. Upton}
\affiliation{Advanced Photon Source, Argonne National Laboratory, Argonne, Illinois 60439, USA}

\author{Vladimir G. Mazurenko}
\affiliation{Ural Federal University, Mira Str. 19, 620002 Ekaterinburg, Russia}

\author{Alexander~A. Tsirlin}
\email{altsirlin@gmail.com}
\affiliation{Experimental Physics VI, Center for Electronic Correlations and Magnetism, Institute of Physics, University of Augsburg, 86135 Augsburg, Germany}
\affiliation{Ural Federal University, Mira Str. 19, 620002 Ekaterinburg, Russia}


\begin{abstract}
Crystal structure, electronic state of Ir$^{4+}$, and magnetic properties of the antifluorite compound K$_2$IrBr$_6$ are studied using high-resolution synchrotron x-ray diffraction, resonant inelastic x-ray scattering (RIXS), thermodynamic and transport measurements, and \textit{ab initio} calculations. The crystal symmetry is reduced from cubic at room temperature to tetragonal below 170\,K and eventually to monoclinic below 122\,K. These changes are tracked by the evolution of the non-cubic crystal-field splitting $\Delta$ measured by RIXS. Non-monotonic changes in $\Delta$ are ascribed to the competing effects of the tilt, rotation, and deformation of the IrBr$_6$ octahedra as well as tetragonal strain on the electronic levels of Ir$^{4+}$. The N\'eel temperature of $T_N=11.9$\,K exceeds that of the isostructural K$_2$IrCl$_6$, and the magnitude of frustration on the fcc spin lattice decreases. We argue that the replacement of Cl by Br weakens electronic correlations and enhances magnetic couplings.
\end{abstract}

\maketitle


\section{Introduction}
Spin-orbit coupling can have major effect on the electronic structure and magnetism of $5d$ transition metals~\cite{krempa2014,rau2016}. The extent of spin-orbit physics depends on whether the orbital moment of a $5d$ ion is fully or partially quenched -- an effect that hinges upon details of crystal-field levels and local environment of the transition-metal atom. For example, electronic state of the $5d^5$ Ir$^{4+}$ ion can change from mundane \mbox{spin-$\frac32$} in square-planar IrO$_4$ complexes~\cite{kanungo2015,ming2017} to a more exotic $j_{\rm eff}=\frac12$ in the regular IrO$_6$ octahedral environment, which has interesting implications for spin-liquid physics and unusual metallic states~\cite{winter2017,cao2018,bertinshaw2019}. 

Experimental studies of iridates suggest that even minor deviations from the ideal symmetry of an IrO$_6$ octahedron may cause appreciable non-cubic crystal-field splittings $\Delta$. Many of the compounds studied to date -- including honeycomb iridates~\cite{gretarsson2013}, double perovskites~\cite{aczel2019,revelli2019}, and fluorides~\cite{rossi2017} -- show the nearly constant $\Delta$ of about 150\,meV~\footnote{Note that Ref.~\cite{aczel2019} refers to $\Delta$ as the splitting between the RIXS peaks, $\Delta_{\rm exp}$. In contrast, our $\Delta$ is the actual splitting between the $t_{2g}$ levels, and $\Delta=\frac32\Delta_{\rm exp}$ in the limit of $\Delta\ll\lambda$, where $\lambda\simeq 0.5$\,eV is the typical spin-orbit coupling in Ir$^{4+}$ compounds.}. This similarity goes across different structure types, with the $\Delta$ value being remarkably insensitive to the structural motif or crystallographic symmetry. For example, similar values of $\Delta$ were reported for monoclinic Sr$_2$CeIrO$_6$ and nominally cubic Ba$_2$CeIrO$_6$~\cite{aczel2019,revelli2019}. This raises a natural question of which structural parameters determine $\Delta$ in the limit of weak local distortions, and whether $\Delta$ values below 150\,meV can be reached in a real Ir$^{4+}$ material.

Here, we shed light on this problem by studying K$_2$IrBr$_6$ and report $\Delta\simeq 50$\,meV, the lowest value of a non-cubic crystal-field splitting achieved so far in an Ir$^{4+}$ compound. We also track the influence of different structural distortions on this parameter by monitoring $\Delta$ upon cooling whilst the compound progressively reduces its symmetry via two consecutive structural phase transitions. K$_2$IrBr$_6$ belongs to the family of Ir$^{4+}$ antifluorites, material realizations of frustrated spin-$\frac12$ fcc antiferromagnets~\cite{lines1963,khan2019,schick2020}, and we also investigate how symmetry lowering changes the strength of electronic correlations and magnetic couplings as well as the magnitude of magnetic frustration in this family of compounds. {\cred Our results are complementary to the very recent study of the Ir$^{4+}$ antifluorites~\cite{plessis2020} where similarly low values of $\Delta$ were reported for K$_2$IrBr$_6$, K$_2$IrCl$_6$, and other compounds of the same family, but temperature evolution of $\Delta$ and its interplay with the structural parameters were not explored.}%

The manuscript is organized as follows. Sec.~\ref{sec:methods} contains methodological details. In Sec.~\ref{sec:structure}, we report crystal structure of K$_2$IrBr$_6$ as a function of temperature, analyze individual distortions, and discuss phonon instabilities that lead to consecutive structural phase transitions in this compound. Sec.~\ref{sec:rixs} describes the electronic state of Ir$^{4+}$ studied by resonant inelastic x-ray scattering and \textit{ab initio} calculations across the different polymorphs. Sec.~\ref{sec:magnetism} contains the results of thermodynamic measurements and \textit{ab initio} calculations of exchange couplings that both probe low-temperature magnetism of K$_2$IrBr$_6$. Sec.~\ref{sec:discussion} concludes the manuscript with a brief discussion and summary.


\section{Methods}
\label{sec:methods}

\begin{figure}
	\centering
	\includegraphics[scale=0.5]{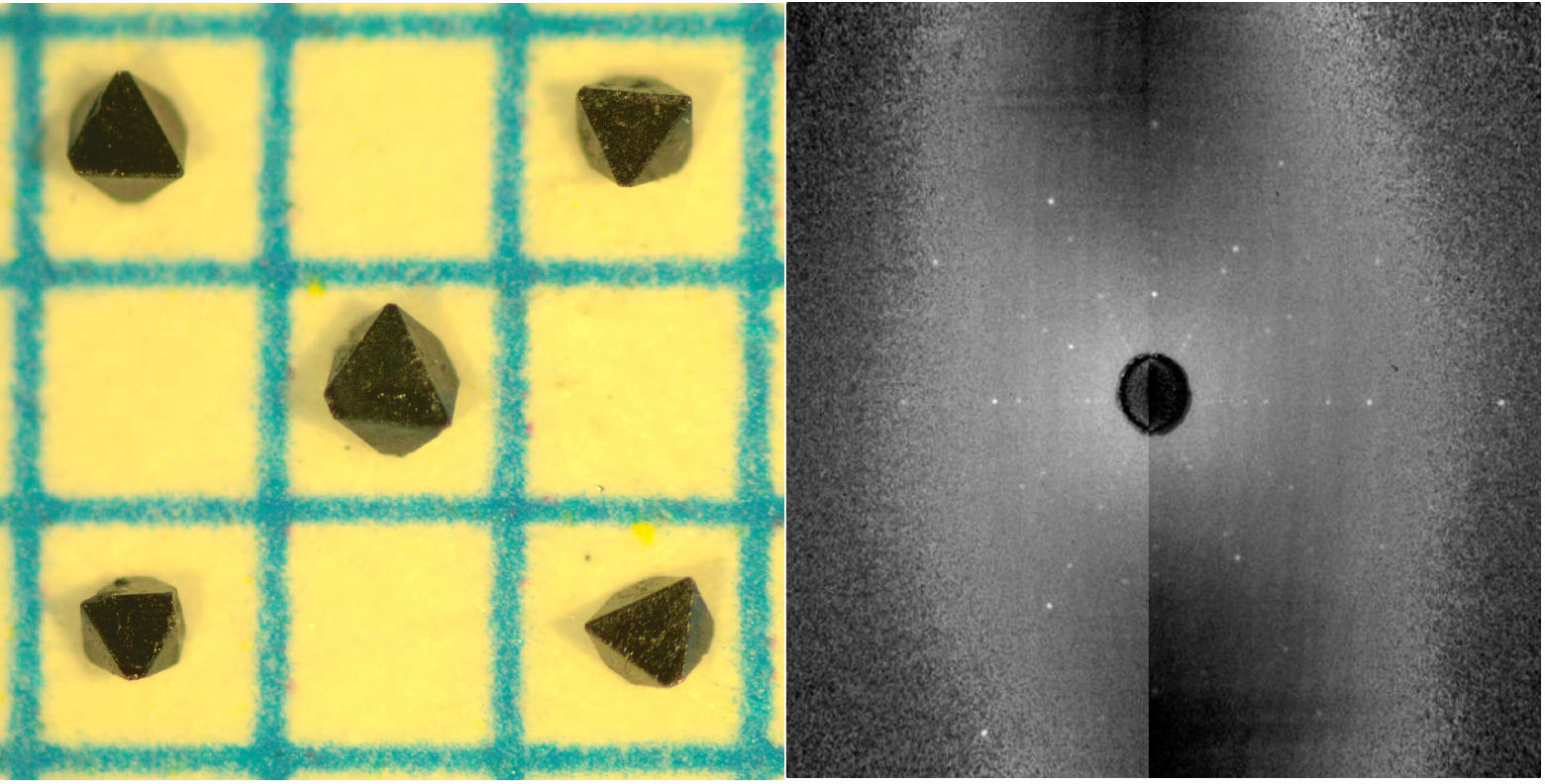}
	\caption{Image of selected crystals on a mm graph paper and x-ray Laue diffraction pattern of a K$_2$IrBr$_6$ single crystal.}\label{Fig.1}
\end{figure}

K$_2$IrBr$_6$ powder was obtained from ChemPUR GmbH and Alfa Aesar (Ir 25.4\% min). Its phase purity was confirmed by room-temperatures powder x-ray diffraction (XRD) data collected at the Rigaku MiniFlex diffractometer (Cu $K_{\rm \alpha}$ radiation). Single crystals were grown from the solution prepared using this powder. A nearly saturated and slightly acidic (${\rm pH}\sim 2$) aqueous solution of K$_2$IrBr$_6$ was placed into a glass beaker and kept inside a box furnace with temperature maintained at 60\,$^\circ$C, for the nucleation of the crystals by slow evaporation of water from the solution. After 4 days many small crystals with the typical size of about (0.7$\times$0.6$\times$0.4) mm$^3$ were found at the bottom of the glass beaker (Fig.~\ref{Fig.1}). X-ray Laue diffraction experiment confirmed high crystallinity of these samples. The back scattered Laue diffraction pattern with x-ray beam parallel to the normal of the crystal face exhibits three-fold symmetry of the reciprocal space (Fig.~\ref{Fig.1}), and thereby confirms that all naturally occurring crystal faces belong to the family of $\lbrace 111 \rbrace_c$ crystallographic planes, where the subscript indicates plane indices given with respect to the cubic axes of the room-temperature K$_2$IrBr$_6$ structure.

Temperature-dependent crystallographic study was performed on powder, because high resolution was required, and extensive twinning of the crystal could be expected upon symmetry lowering. High-resolution powder XRD was performed at the ID22 beamline of the European Synchtron Radiation Facility (Grenoble, France) using the wavelength $\lambda$ = 0.40001\,\r A at temperatures down to 20\,K stabilized using He flow cryostat. Structure refinement was performed using the \texttt{JANA2006} software~\cite{jana2006}. Additionally, structural phase transitions were probed by a thermal expansion measurement down to 1.78\,K using a highly sensitive capacitive dilatometer~\cite{kuechler2017} operated within Quantum Design PPMS. The data were collected in both heating and cooling cycles with a temperature sweep rate of 0.2\,K/min.

Resonant inelastic x-ray scattering (RIXS) measurements were performed on the single crystal of K$_2$IrBr$_6$ using the MERIX spectrometer on beamline 27-ID of the Advance Photon Source (APS) at Argonne National Laboratory to investigate crystal-field excitations of Ir$^{4+}$. The incident x-ray energy was tuned to the Ir $L_3$ absorption edge at 11.215\,keV. 

Temperature and field dependence of the dc magnetization was measured on powder, on individual single crystals, and on a stack of co-aligned single crystals with the total mass of 2.8\,mg using the Quantum Design SQUID-VSM magnetometer (MPMS 3). Pulsed-field magnetization data up to 57\,T were collected in the Dresden High Magnetic Field Laboratory on the powder sample~\cite{tsirlin2009}. The pulsed-field data were then scaled with the magnetization data measured with the SQUID-VSM magnetometer in static fields up to 7\,T. 

Specific heat of the K$_2$IrBr$_6$ single crystal was measured down to 0.5\,K using Quantum Design Physical Properties Measurement System (QD-PPMS) equipped with ${}^{3}$He refrigerator. The conventional two-$\tau$ relaxation method was used. 

The temperature dependence of the dc electrical resistivity was measured using B2987A Electrometer/High Resistance Meter in a commercial PPMS cryostat equipped with a home-made resistivity set-up. Standard two-probe method was employed for the resistivity measurement where two Cu wires were attached to the sample with high-conducting silver paste. The electric field was along the $\langle111\rangle_c$.      

Orbital energies and exchange couplings were assessed by scalar-relativistic density-functional (DFT) band-structure calculations performed in the \texttt{FPLO} code~\cite{fplo} using the $8\times 8\times 6$ k-mesh for the tetragonal and monoclinic structures. Phonon spectra were calculated in \texttt{PHONOPY} via the frozen-phonon method~\cite{phonopy}. The forces for the phonon calculations were obtained from DFT and DFT+$U$+SO calculations in \texttt{VASP}~\cite{vasp1,vasp2} using the $2\times 2\times 1$ supercell and 0.01\,\r A atomic displacements. Perdew-Burke-Ernzerhof flavor of the exchange-correlation potential~\cite{pbe96} was chosen for all calculations. DFT+$U$+SO calculations utilized double-counting correction in the atomic limit and the Hund's coupling parameter $J=0.3$\,eV. The choice of the on-site Coulomb repulsion parameter $U$ is further discussed in Sec.~\ref{sec:correlations}.

\section{Structural transformations}
\label{sec:structure}

\begin{figure}
\includegraphics{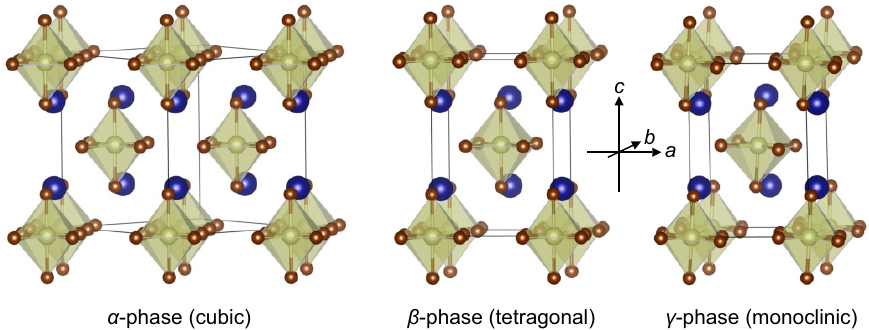}
\caption{\label{fig:structure}
Crystal structures of the K$_2$IrBr$_6$ polymorphs.
}
\end{figure}
\begin{figure}
	\centering
	\includegraphics[width=8.5cm]{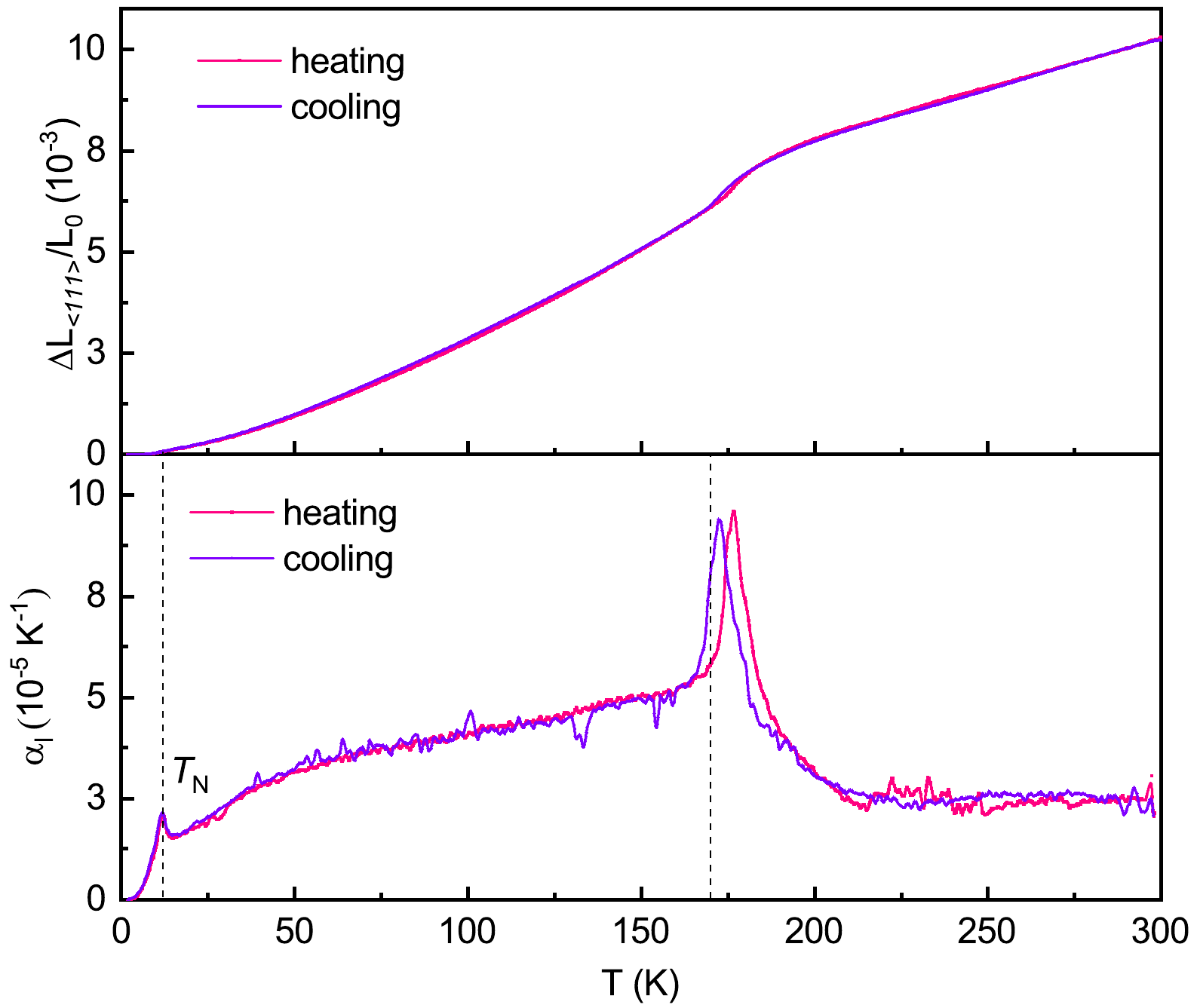}
	\caption{Relative length change $\Delta L/L$ of the K$_2$IrBr$_6$ single crystal measured along the $\langle 111\rangle_c$ crystallographic direction, and the corresponding linear thermal expansion coefficient, $\alpha_l=1/L_0[d(\Delta L)/dT]$, as a function of temperature. Hysteretic behavior is clearly seen at the $\alpha-\beta$ transition around 170\,K. A weaker anomaly around 12\,K indicates the onset of magnetic long-range order.
	\label{fig:te}
	}
\end{figure}

Antifluorite-type crystal structure of K$_2$IrBr$_6$ features isolated IrBr$_6$ octahedra arranged on the sites on an fcc lattice and separated by K atoms (Fig.~\ref{fig:structure}). The idealized structure is cubic, but rotations, tilts, and deformations of the octahedra may lead to different types of symmetry lowering.

Early temperature-dependent studies reported two phase transitions in K$_2$IrBr$_6$ at 182\,K and 13\,K from differential thermal analysis and calorimetry~\cite{rossler1977}. No diffraction experiments were performed. Our thermal expansion (Fig.~\ref{fig:te}) and specific heat (Fig.~\ref{fig:heat}) data confirm the transitions around 170\,K and 12\,K, respectively. Magnetic susceptibility changes upon the latter transition (Fig.~\ref{fig:chi}) suggesting magnetic ordering rather than structural transformation as the primary origin of this anomaly. High-resolution XRD additionally shows another structural phase transition around 120\,K that lacks any signatures in thermodynamic measurements. Cubic symmetry of the room-temperature crystal structure ($\alpha$-K$_2$IrBr$_6$) is reduced to tetragonal below 170\,K ($\beta$-K$_2$IrBr$_6$) and to monoclinic below 122\,K ($\gamma$-K$_2$IrBr$_6$). {\cred This sequence of structural phase transitions is consistent with the very recent neutron diffraction study~\cite{plessis2020}.}

\subsection{Cubic $\alpha$-phase}

\begin{figure}
	\centering
	\includegraphics[scale=0.55]{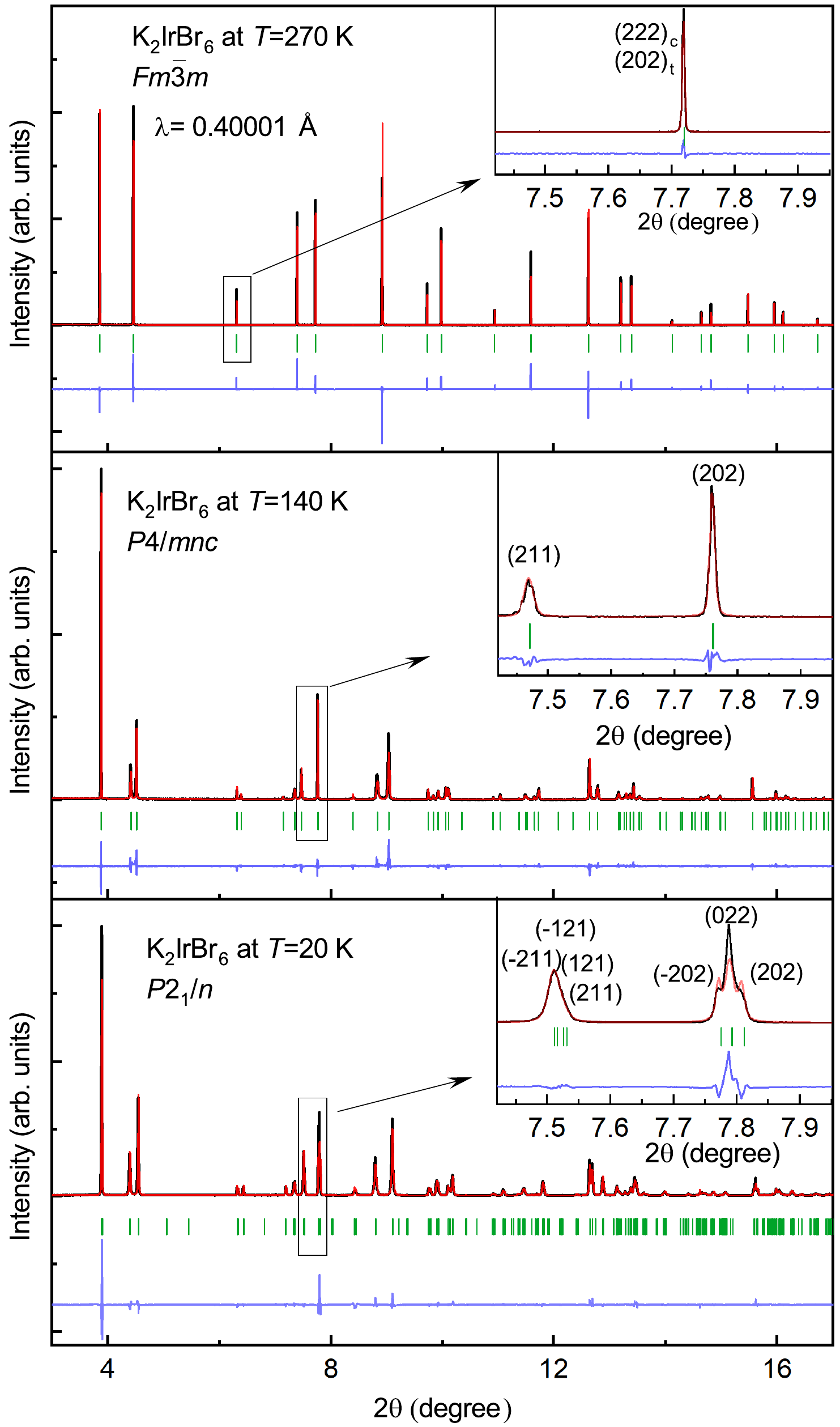}
	\caption{Structure refinements for $\alpha$-K$_2$IrBr$_6$ at 270\,K, $\beta$-K$_2$IrBr$_6$ at 140\,K, and $\gamma$-K$_2$IrBr$_6$ at 20\,K. The solid red lines are the calculated patterns. The insets show the temperature evolution of the cubic (222) reflection, which at 20\,K exhibits a clear monoclinic splitting, whereas the (211) reflection is forbidden in the fcc structure and appears only in the $\beta$-phase.}
 \label{fig:patterns}
\end{figure}

At 270\,K, sharp and non-split reflections measured by high-resolution synchrotron XRD (Fig.~\ref{fig:patterns}) confirm cubic symmetry of the crystal structure. The refined lattice parameter $a=10.29266(2)$\,\r A is consistent with the reported room-temperature value of $a=10.298(5)$\,\r A~\cite{gubanov2002}. Br atoms are at $(x,0,0)$ with the Ir--Br distance of 2.466(2)\,\r A, about 0.15\,\r A longer than in isostructural K$_2$IrCl$_6$~\cite{khan2019}.

\begin{figure}
\includegraphics{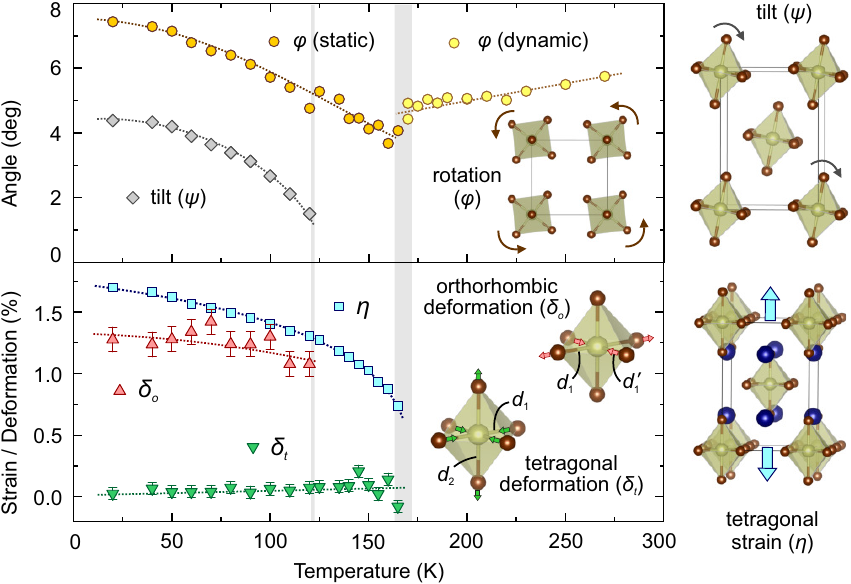}
\caption{\label{fig:distortion}
Distortions of the K$_2$IrBr$_6$ structure. Upper panel: rotation ($\varphi$) and tilt ($\psi$) of the IrBr$_6$ octahedra. Lower panel: tetragonal ($\delta_t$) and orthorhombic ($\delta_o$) deformations of the octahedra, as well as tetragonal strain ($\eta$). See text for the definitions of $\delta_t$, $\delta_o$, and $\eta$. The lines are guide for the eye.
}
\end{figure}

\begin{figure}
	\centering
	\includegraphics[width=8.5cm]{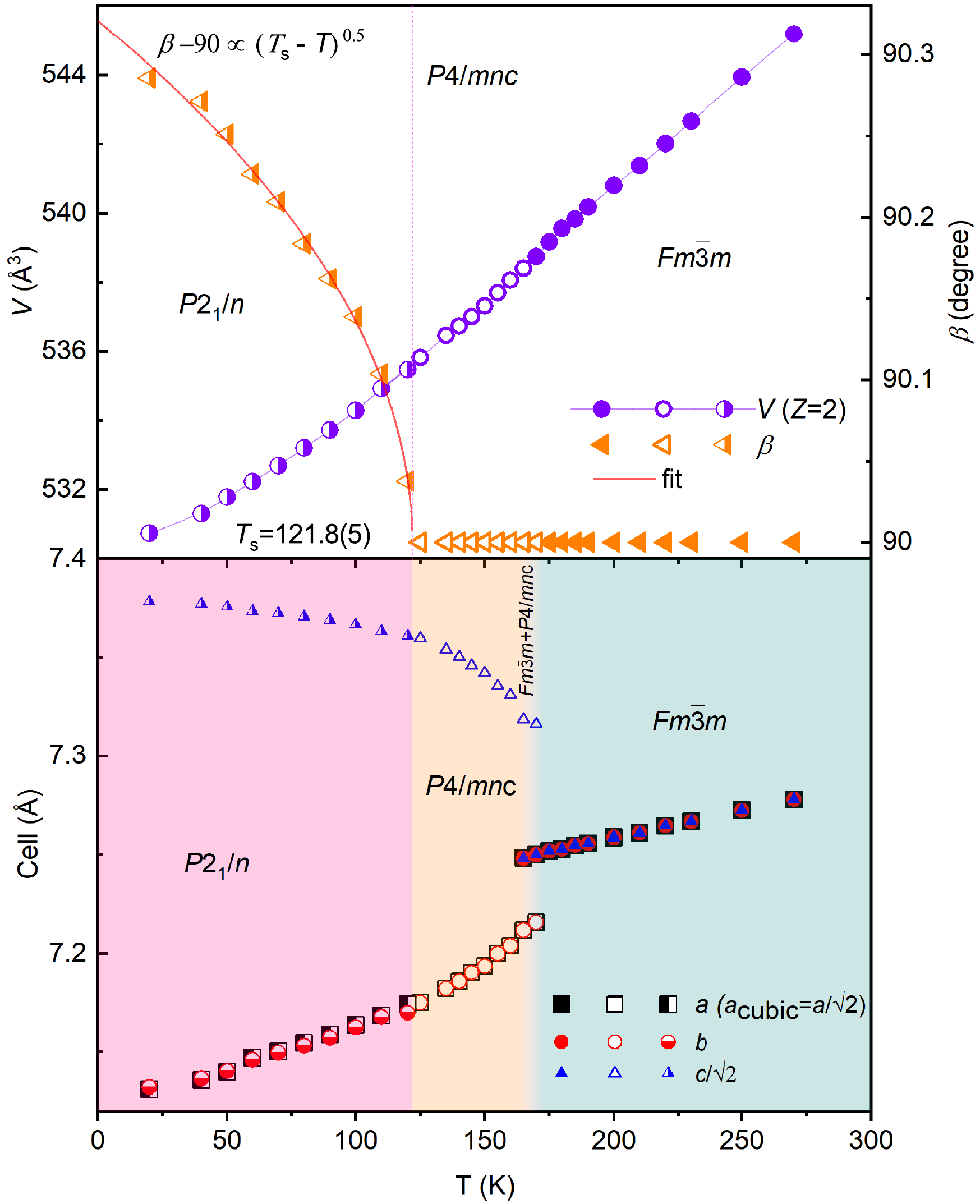}
	\caption{Lattice parameters of K$_2$IrBr$_6$. Upper panel: unit cell volume (two formula units) and monoclinic angle $\beta$ along with the power-law fit described in the text. Lower panel: $a$, $b$, and $c/\sqrt{2}$ given in the setting of $\beta$- and $\gamma$-K$_2$IrBr$_6$. \label{fig:lattice}
	}
\end{figure}

The inspection of atomic displacement parameters (ADPs) at 270\,K revealed anomalously high values for K and Br (Table~\ref{tab:structure}). In the case of Br, displacement ellipsoid is stretched in the direction perpendicular to the Ir--Br bond ($U_{22}=U_{33}\gg U_{11}$) and indicates disordered rotations of the IrBr$_6$ octahedra. Similar features have been reported in isostructural K$_2$IrCl$_6$~\cite{khan2019}. 

The rotation angle $\varphi$ is gauged using the transverse component of the Br ADP. With $U_{22}=U_{33}=\langle(\Delta r)^2\rangle$, one estimates $\varphi={\rm tan}^{-1}(\sqrt{U_{22}}/d)$, where $d$ is the Ir--Br distance. Temperature evolution of $\varphi$ (Fig.~\ref{fig:distortion}) reveals that the rotations are gradually suppressed upon cooling, but remain sizable even at 170\,K where they condense causing the $\alpha-\beta$ transition.

\subsection{Tetragonal $\beta$-phase}

Below 170\,K, additional reflections incompatible with the face-centered cubic structure appeared in the XRD pattern (Fig.~\ref{fig:patterns}). They could be indexed in a primitive tetragonal unit cell with $a_{\rm tet}=a_{\rm cub}/\sqrt 2$ and $c_{\rm tet}=c_{\rm cub}$. The crystal structure was refined in the space group $P4/mnc$ determined previously for Rb$_2$TeI$_6$~\cite{abriel1982} and several other distorted antifluorite compounds~\cite{abriel1984,abrahams1984}. K and Ir positions remain fully constrained by symmetry, whereas the Br position splits into two, resulting in two types of distortions: i) cooperative rotations of the IrBr$_6$ octahedra in the $ab$ plane; ii) tetragonal deformation of the octahedra. Our structure refinement shows that the distortion is dominated by the former effect. The rotation angle $\varphi$ lies in the range of $4-5^{\circ}$ (Fig.~\ref{fig:distortion}) suggesting that dynamic rotations in the $\alpha$-phase become static below the transition. Tetragonal deformation is gauged by $\delta_t=(d_2-d_1)/(d_1+d_2)$, where $d_1$ and $d_2$ are the in-plane and out-of-plane distances in the IrBr$_6$ octahedron, respectively. This deformation remains negligible throughout the stability range of the $\beta$-phase.

\begin{figure}
\includegraphics[width=8.6cm]{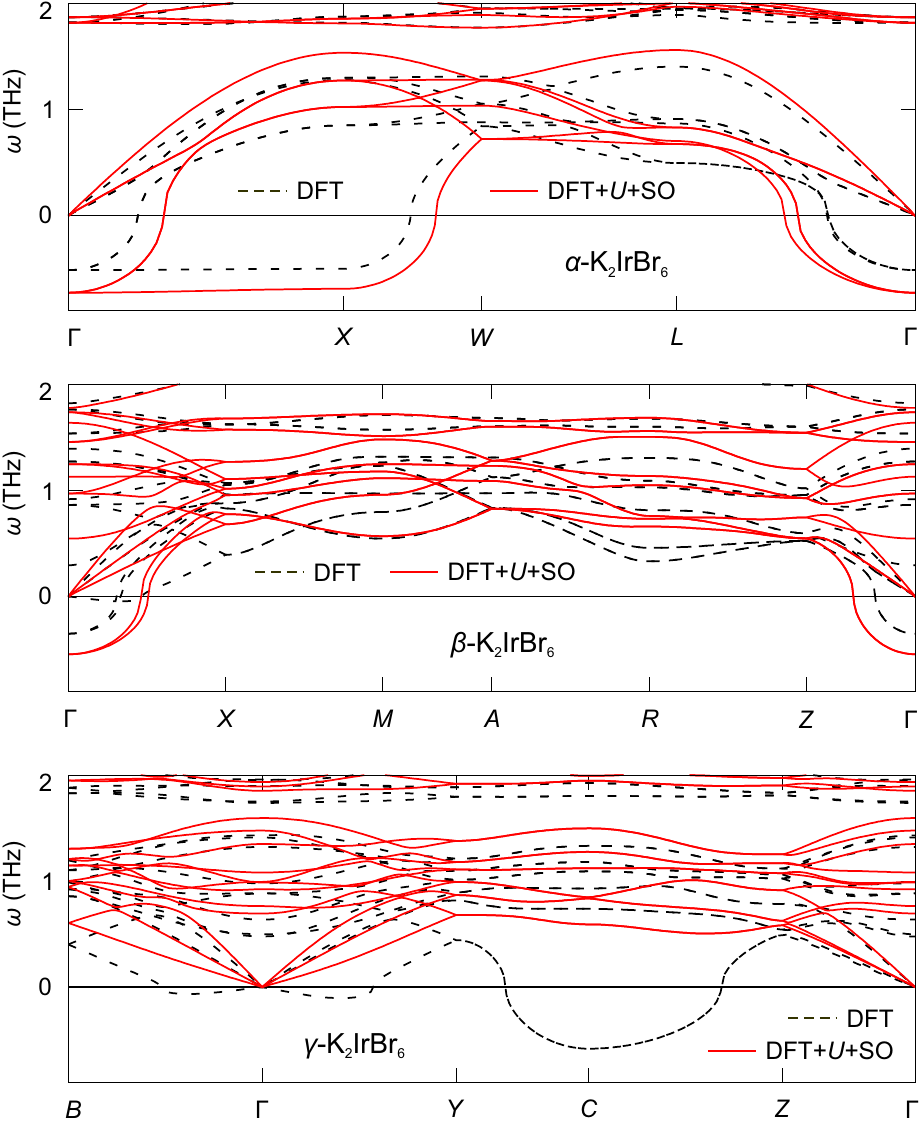}
\caption{Calculated phonon spectra for different polymorphs of K$_2$IrBr$_6$. The dashed black and solid red lines are phonon branches obtained within DFT and DFT+$U$+SO, respectively. }
\label{fig:phonon}
\end{figure}
First-order nature of the $\alpha-\beta$ transition can be inferred from the hysteretic behavior seen in the thermal expansion measurement (Fig.~\ref{fig:te}) and from phase coexistence revealed by XRD. At several temperatures around the transition (Fig.~\ref{fig:lattice}), both cubic and tetragonal phases were present in the sample, with the fraction of the $\beta$-phase decreasing from 76\,\% at 165\,K to 26\,\% at 170\,K upon heating~\cite{suppl}. 

The cubic-to-tetragonal transition in antifluorite compounds is caused by soft phonon modes observed at the zone center ($\Gamma$) and zone boundary ($X$)~\cite{lynn1978,mintz1979}. Indeed, our phonon calculations for $\alpha$-K$_2$IrBr$_6$ reveal an imaginary phonon band along $\Gamma - X$ (Fig.~\ref{fig:phonon}). This instability present in both DFT and DFT+$U$+SO calculations is a rotary mode that causes dynamic rotations of the IrBr$_6$ octahedra. It originates from a size mismatch between the octahedra and K atoms~\cite{sutton1983}.

The condensation of a soft phonon mode normally leads to a second-order phase transition, as reported in K$_2$OsCl$_6$~\cite{novotny1977,armstrong1978}, K$_2$ReCl$_6$~\cite{willemsen1977}, and K$_2$TeBr$_6$~\cite{abrahams1984}. In contrast, our data reveal strong signatures of a first-order transition, even though it follows group-subgroup relation, and octahedral rotations serve as the primary order parameter. In K$_2$IrBr$_6$, first-order character of the transition is likely caused by a strong coupling to the tetragonal strain~\cite{cowley1980}, $\eta=(a_{\rm tetr}-c_{\rm tetr}/\sqrt 2)/(a_{\rm tetr}+c_{\rm tetr}/\sqrt 2)$, that appears abruptly at the transition ($\eta_{\rm 170\,K}=0.69$\%) and systematically increases down to low temperatures ($\eta_{\rm 20\,K}=1.72$\%), as shown in Fig.~\ref{fig:distortion}. In contrast, K$_2$TeBr$_6$ with the second-order $\alpha-\beta$ transition features $\eta\simeq 0$ at the transition, followed by a linear increase of $\eta$ to 0.39\% within a narrow temperature range below the transition, and an eventual decrease down to $\eta=0.33$\% at 20\,K~\cite{abrahams1984}. 

\begin{table*}
\caption{Structural parameters of $\alpha$-K$_2$IrBr$_6$ at 270 K, $\beta$-K$_2$IrBr$_6$ at 140 K, and $\gamma$-K$_2$IrBr$_6$ at 20 K. The error bars are from the Rietveld refinement. Atomic displacement parameters $U_{\rm iso}$ are given in \r A$^2$. All the crystallographic sites are fully occupied. {\cred The $R_I$ and $R_p$ are refinement residuals for the peak intensities and peak profile, respectively~\cite{suppl}.}}\label{tab:structure}
\centering
\newcommand*{\TitleParbox}[1]{\parbox[c]{1.8cm}{\raggedright #1}}
\begin{tabular*}{1\textwidth}{@{\extracolsep{\fill}}c c c c}
\hline
\hline  
$T(K)$ & 270 K & 140 K & 20 K\\ 
\hline 
Space group & $Fm$\={3}$m$ & $P4/mnc$ & $P2_{1}/n$ \\ 
$a$({\AA}) & 10.29266(2) & 7.18558(8) & 7.13121(10)\\
$b$({\AA}) & 10.29266(2) & 7.18558(8) & 7.13240(10)\\
$c$({\AA}) & 10.29266(2) & 10.39521(14) & 10.43467(15) \\ 
$\beta$ (deg) & 90 & 90 & 90.2857(5)\\
 
$R_{I}/R_{P}$ & 0.0765/0.1663 & 0.0699/0.1720 & 0.0303/0.0923 \\ 
\hline 
Ir & 4$a$ (0,0,0)& 2$a$ (0,0,0) & 2$a$ (0,0,0) \\ 

& $U_{\rm iso}=0.0166(1)$ & $U_{\rm iso}=0.0115(4)$ & $U_{\rm iso}=0.0033(4)$ \\ 

\hline 
K & 8$c$ ($\frac{1}{4}$,$\frac{1}{4}$,$\frac{1}{4}$) & 4$d$ (0,$\frac{1}{2}$,$\frac{1}{4}$) & 4$e$ ($x$,$y$,$z$) \\ 

 & & & $x=-0.0077(8)$ \\ 

 &  & & $y=0.4941(14)$ \\ 
 &  & & $z=0.2583(6)$ \\ 

 & $U_{\rm iso}=0.0409(6)$ & $U_{\rm iso}=0.0317(15)$ & $U_{\rm iso}=0.0143(16)$ \\ 
\hline 
Br(1) & 24$e$ $(x,0,0)$ & 4$e$ $(0,0,z)$ & 4$e$ $(x,y,z)$ \\ 

 & $x=0.2373(1)$ & & $x=0.0263(3)$  \\ 
 & &  & $y=-0.0030(7)$  \\ 
 & & $z=0.2376(2)$ & $z=0.2361(1)$  \\ 
 & $U_{\rm iso}=0.0346(4)$ & $U_{\rm iso}=0.0295(4)$ & $U_{\rm iso}=0.0087(7)$ \\
\hline  
Br(2) & & 8$h$ $(x,y,0)$ & 4$e$ $(x,y,z)$ \\ 
      & & $x=0.2231(2)$  & $x=0.2803(4)$  \\ 
      & & $y=0.2607(2)$  & $y=-0.2106(3)$  \\ 
      & & & $z=-0.0125(2)$  \\ 
      & & $U_{\rm iso}=0.0295(4)$ & $U_{\rm iso}=0.0136(6)$ \\
\hline 
 Br(3) &  &  & 4$e$ $(x,y,z)$ \\ 
 &  &  & $x=0.2156(3)$ \\ 
 &  &   & $y=0.2638(3)$ \\ 
  &  &   & $z=-0.0102(2)$ \\ 
 & &  & $U_{\rm iso}=0.0136(6)$ \\
\hline
\hline 
\end{tabular*}
\end{table*}

\subsection{Monoclinic $\gamma$-phase}

The second structural phase transition in K$_2$IrBr$_6$ has no fingerprints in thermal expansion or specific heat, but can be clearly seen in high-resolution XRD data where (202) reflection of the $\beta$-phase splits into three peaks indicative of a monoclinic distortion (Fig.~\ref{fig:patterns}). {\cred The absence of this splitting above 120\,K attests tetragonal symmetry of the $\beta$-phase and suggests that an additional phase transition should take place at the temperature where peak splitting appears. This transition is further evidenced by the RIXS data that probe local environment of Ir$^{4+}$ and its non-cubic crystal-field splitting (see Sec.~\ref{sec:rixs}). In the case of K$_2$PtBr$_6$, a spectroscopic probe (nuclear quadrupolar resonance) was also shown to be most sensitive to distortions beyond tetragonal symmetry in antifluorite hexahalide compounds~\cite{vandriel1972}.}%

The monoclinic angle increases upon cooling. Its temperature dependence can be approximated by $(\beta-90^\circ)\propto (T_{s}-T)^{0.5}$ and returns the transition temperature of $T_s=121.8(5)$\,K. The power-law evolution of $\beta$ corresponds to a second-order phase transition in agreement with the absence of a volume change or hysteresis around $T_s$ (Fig.~\ref{fig:lattice}).

Symmetry lowering toward $P2_1/n$ allows several types of distortions, most notably the tilt of the IrBr$_6$ octahedra relative to the $c$-axis and the orthorhombic deformation of the octahedra (Figs.~\ref{fig:structure} and~\ref{fig:distortion}). The $\beta-\gamma$ transition is driven by a phonon instability of the $\beta$-phase where imaginary phonon modes are found around the $\Gamma$-point (Fig.~\ref{fig:phonon}). {\cred Although metrically tetragonal, $\beta$-phase still features a significant amount of dynamic disorder, as seen from the increased ADPs of K and Br atoms (Table~\ref{tab:structure}).} In contrast, the $\gamma$-phase {\cred shows much lower ADPs}. This structure is dynamically stable, with no imaginary phonon frequencies in DFT+$U$+SO. Imaginary phonons are still obtained in the DFT calculation, but they are a drawback of neglecting Coulomb correlations and spin-orbit coupling, similar to $\alpha$-RuCl$_3$ where the spurious dimerized state is stabilized on the DFT level~\cite{widmann2019}.

Detailed analysis of the $\gamma$-phase structure shows that orthorhombic strain (the difference between $a$ and $b$) remains negligible, whereas tetragonal strain $\eta$ increases steadily upon cooling (Fig.~\ref{fig:distortion}). In-plane octahedral rotations $\varphi$ increase too and reach nearly $7.5^{\circ}$ at 20\,K. Additionally, the octahedral tilt $\psi$ and the orthorhombic deformation $\delta_o$ develop below the $\beta-\gamma$ transition (Fig.~\ref{fig:distortion}). Here, we define $\delta_o=(d_1-d_1')/(d_1+d_1')$, where $d_1$ and $d_1'$ are the in-plane distances in the IrBr$_6$ octahedron. The orthorhombic deformation reaches about 1.3\% at 20\,K, whereas tetragonal deformation $\delta_t$ remains small. In contrast, crystal structures of Ir$^{4+}$ oxides are often dominated by the tetragonal deformation, such as $\delta_t=-1.9$\% in Sr$_2$IrO$_4$~\cite{huang1994} and post-perovskite CaIrO$_3$~\cite{nakatsuka2015}.

\section{Electronic Structure}
\label{sec:rixs}

\subsection{Crystal-field splitting}

\begin{figure}
	\centering
	\includegraphics[scale=0.5]{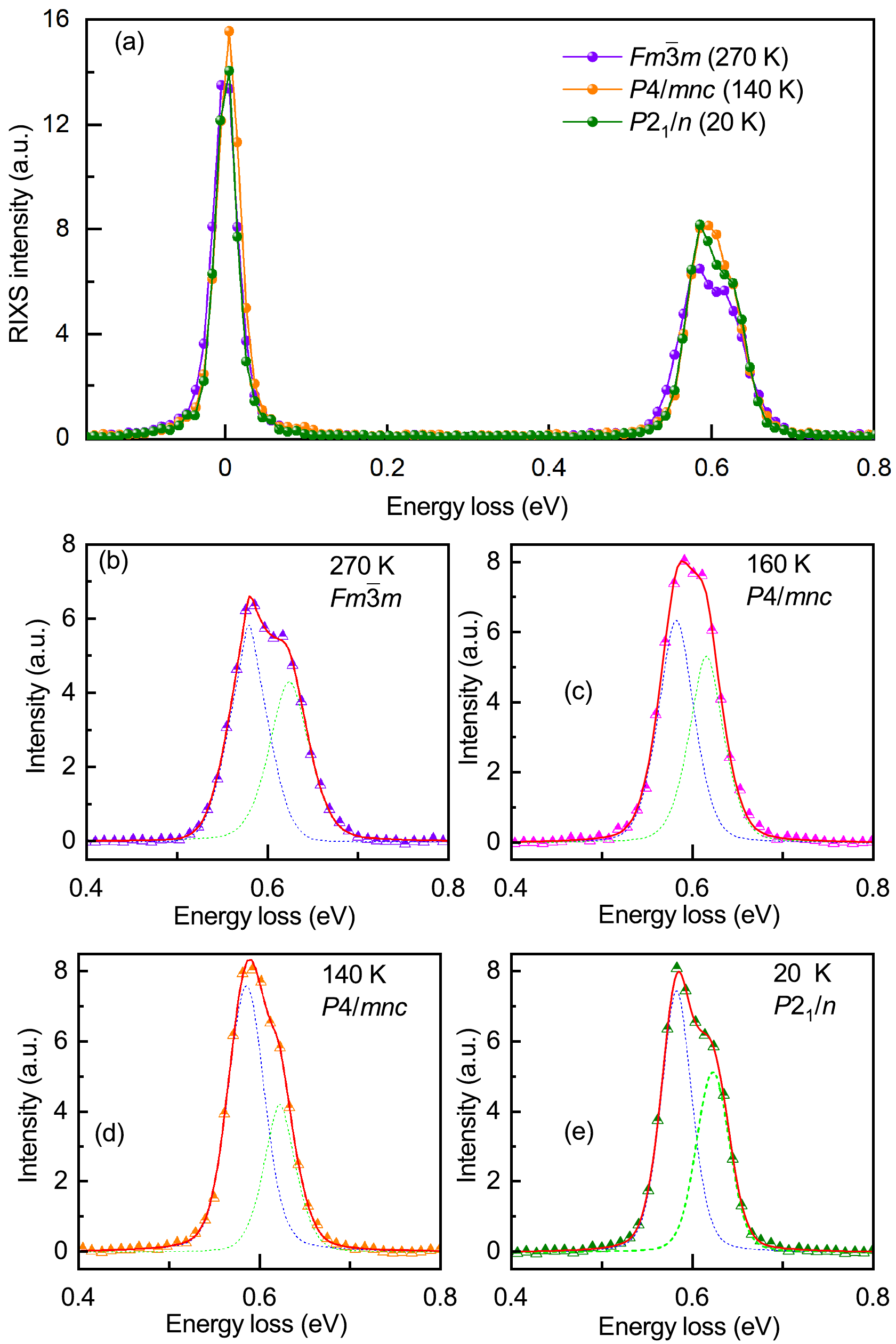}
	\caption{(a) RIXS spectra measured at the Ir $L_3$ edge ($E_i=11.215$\,keV) for different polymorphs. The peak at 0\,eV corresponds to the strong elastic line, whereas the peaks at $0.5-0.7$\,eV are inelastic lines due to the low-lying excitations. The double-peak structure of this feature manifests a splitting of the $j=\frac32$ states. (b-e) Enlarged view of the inelastic lines at selected temperatures. The solid lines are fits with two pseudo-Voigt functions. The fit to the 160\,K spectrum in panel (c) returns the lowest crystal-field splitting of $\Delta=49$\,meV.} \label{fig:spectra}
\end{figure} 
\begin{figure}
	\centering
	\includegraphics[scale=0.95]{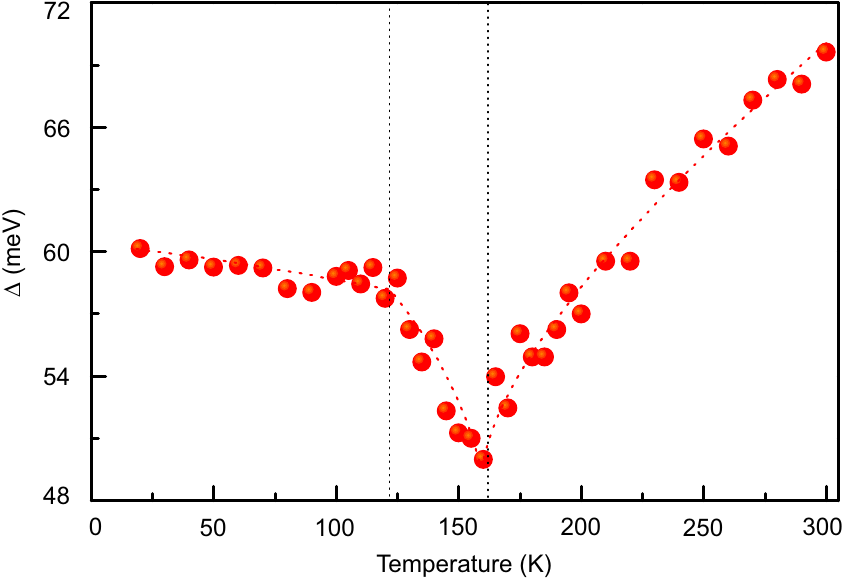}
	\caption{Temperature dependence of the noncubic crystal-field splitting $\Delta$ estimated from the two-peak pseudo-Voigt fits to the RIXS spectra. Dotted lines are guide to the eye.} \label{fig:delta}
\end{figure}

Our structural analysis in Sec.~\ref{sec:structure} reveals the non-cubic crystal environment of Ir$^{4+}$ in K$_2$IrBr$_6$. Indeed, even in the nominally cubic $\alpha$-phase rotations of the IrBr$_6$ octahedra may lower the symmetry locally. To assess the effect of these structural distortions onto the electronic levels of Ir$^{4+}$, we performed RIXS measurements at the Ir $L_3$ edge. 

Under ideal cubic symmetry and in the presence of spin-orbit coupling, the $t_{2g}$ levels of Ir$^{4+}$ transform into the lower-lying $j_{\rm eff}=\frac32$ and higher-lying $j_{\rm eff}=\frac12$ states. The lowest-energy excitation is then of the $\frac32\rightarrow\frac12$ nature and appears at $\frac32\lambda$, where $\lambda$ is the spin-orbit coupling constant. K$_2$IrBr$_6$ reveals this excitation in the form of a weakly split peak centered around 0.6\,eV (Fig.~\ref{fig:spectra}). The separation into two peaks manifests residual non-cubic crystal-field splitting in the Ir $t_{2g}$ shell. 

\begin{figure}
\includegraphics{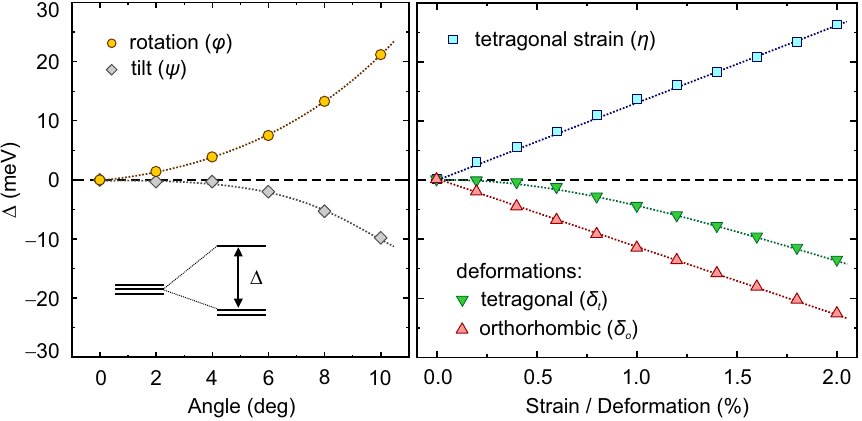}
\caption{\label{fig:delta-cal}
Noncubic crystal-field splitting $\Delta$ calculated for different types of local distortions introduced in Fig.~\ref{fig:distortion}. The inset shows the level scheme corresponding to $\Delta>0$. 
}
\end{figure}

For a quantitative analysis, the spectral feature at 0.6\,eV was fitted to the sum of two pseudo-Voigt functions with the peak positions $\hslash\omega_1$ and $\hslash\omega_2$. Assuming $\Delta\ll\lambda$, one finds $(\hbar\omega_1+\hbar\omega_2)/2=\frac32\lambda$ and $\Delta=\frac32(\hbar\omega_2-\hbar\omega_1)$~\cite{moretti2014} {\cred (note the difference to Refs.~\onlinecite{aczel2019,plessis2020} where $\Delta$ is taken as the separation between $\hbar\omega_1$ and $\hbar\omega_2$ without the $\frac32$ pre-factor).} We thus estimate $\lambda=0.40$\,eV, which is lower than $\lambda=0.45-0.47$\,eV in Ir$^{4+}$ oxides~\cite{revelli2019,aczel2019} and $\lambda=0.57$\,eV in Ir$^{4+}$ fluorides~\cite{rossi2017}. This reflects the increased covalency of the Ir--Br interactions, as further discussed in Sec.~\ref{sec:correlations}.

Weak splitting of the $t_{2g}$ levels leads to a slight mixing of the $j_{\rm eff}=\frac12$ and $j_{\rm eff}=\frac32$ states. The ground-state wave function of Ir$^{4+}$ in the $\vert j, j_z)$ basis takes the form 
\begin{equation}
\left\vert 0 \right\rangle = \cos\theta \left\vert \dfrac{1}{2},\dfrac{1}{2}\right\rangle+\sin\theta \left\vert \dfrac{3}{2},\dfrac{1}{2}\right\rangle, \label{Eq.2}
\end{equation} 
where the mixing angle $\theta$ is given by~\cite{rau2016}
\begin{equation}
\tan2\theta=\frac{4\sqrt 2\Delta}{2\Delta+9\lambda}. \label{Eq.4}
\end{equation}     
We estimate $\cos\theta=0.9993$ using $\Delta\simeq 49$\,meV at 160\,K. This indicates more than 99.9\% contribution of the $j_{\rm eff}=\frac12$ state to the ground-state wavefunction of Ir$^{4+}$ in K$_2$IrBr$_6$. For comparison, this contribution is 99.4\% in a typical Ir$^{4+}$ oxide with $\Delta\simeq 150$\,meV.

Deviations from the cubic symmetry have a relatively minor influence on the electronic state of Ir$^{4+}$. Nevertheless, it is instructive to track $\Delta$ as a function of temperature and identify main structural effects behind its evolution. Fig.~\ref{fig:delta} shows that $\Delta$ decreases upon cooling within the $\alpha$-phase, increases in the $\beta$-phase, and remains nearly constant in the $\gamma$-phase. To elucidate this non-monotonic evolution, we analyzed the effect of individual structural distortions using DFT calculations~\cite{suppl}. 

Fig.~\ref{fig:delta-cal} reveals several trends that may seem counter-intuitive at first glance. Tetragonal strain ($\eta$), octahedral rotations ($\varphi$), and octahedral tilts ($\psi$) all cause non-cubic crystal-field splittings, even though no distortion is introduced in the IrBr$_6$ octahedra. Tetragonal elongation of the octahedra ($\delta_t>0$) renders $\Delta<0$, which is opposite to the simple electrostatic picture where elongation stabilizes the $d_{yz}$ and $d_{xz}$ orbitals with $\Delta>0$. Finally, orthorhombic deformation ($\delta_o$) has a much stronger effect than the tetragonal one~\footnote{Here, we neglect the weak splitting of $3-4$\,meV that appears between the doubly-degenerate levels upon orthorhombic deformation.}. All these observations suggest that atoms beyond nearest-neighbor bromines, and especially K$^+$ ions that change their positions relative to Br$^-$, contribute to the non-cubic crystal-field splittings in K$_2$IrBr$_6$. Similar effects of distant neighbors were previously reported in the Ir$^{4+}$ oxides~\cite{bogdanov2015}.

The non-monotonic temperature evolution of $\Delta$ can be understood as follows. In the $\alpha$-phase, dynamic rotations of the octahedra are gradually suppressed upon cooling. This lowers $\varphi$ and leads to a monotonic reduction in $\Delta$. The increasing $\Delta$ in the $\beta$-phase is due to the development of tetragonal strain. On the other hand, the constant $\Delta$ value in the $\gamma$-phase should be a result of several competing effects, most notably the orthorhombic deformation $\delta_o$ that compensates the increased rotations and strain. The lowest $\Delta\simeq 49$\,meV is thus reached at the $\alpha-\beta$ transition around $160-170$\,K where octahedral rotations are most strongly suppressed, while tetragonal strain only starts to develop. This puts K$_2$IrBr$_6$ in the nearest proximity of the relativistic $j_{\rm eff}=\frac12$ state compared to all materials reported previously. {\cred K$_2$IrCl$_6$ shows only a slightly larger $\Delta\simeq 72$\,meV at 10\,K~\cite{plessis2020}.}

\subsection{Covalency and correlations}
\label{sec:correlations}
\begin{figure}
	\centering
	\includegraphics[scale=0.45]{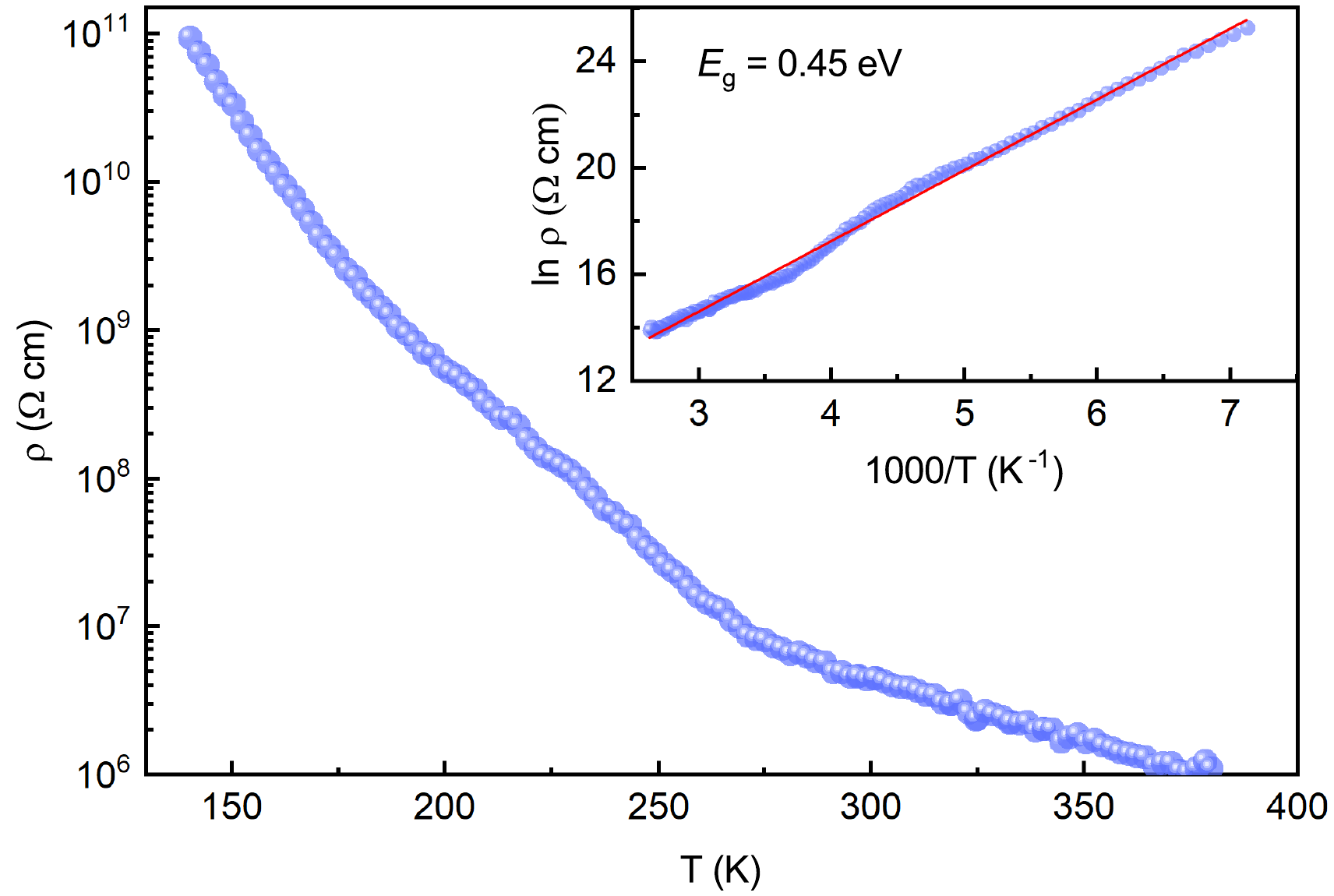}
	\caption{Temperature dependence of zero-field resistivity ($\rho$) measured on a single crystal from 380 K down to 140 K with $y$-axis as log-scale. The resistivity was measured along $\langle111\rangle_c$ with electric field $E\,\|\,\langle 111 \rangle_c$. The inset shows $\ln\rho$ vs $1/T$ and the solid line is the linear fit.} \label{fig:res}
\end{figure} 
The reduced spin-orbit coupling constant $\lambda$ gives first indications of the strong hybridization between the Ir and Br states. This hybridization should screen electronic correlations on the Ir site and thus reduce the band gap. Indeed, our resistivity data reveal a lower band gap compared to isostructural K$_2$IrCl$_6$.

The dc resistivity ($\rho$) of the K$_2$IrBr$_6$ single crystal increases five orders of magnitude upon cooling from 380\,K to 140\,K and exceeds 4\,M$\Omega$ even at room temperature, consistent with the Mott-insulating nature of the compound (Fig.~\ref{fig:res}). The inset of Fig.~\ref{fig:res} shows that the $T$-dependence of $\rho$ follows the activated behavior
\begin{equation}
\rho(T)={\rm exp}\left(\frac{E_g}{2k_B T}\right),\label{Eq.1}
\end{equation}
where $E_g\simeq 0.45$\,eV is the activation energy and $k_{\rm B}$ is the Boltzmann constant. This activation energy is notably smaller than $E_g\simeq 0.7$\,eV in K$_2$IrCl$_6$~\cite{khan2019}. 

The effect of electronic correlations is gauged by comparing experimental activation energies $E_g$ with the DFT+$U$+SO band gap obtained for different values of $U$. Best agreement found at $U=1.8$\,eV for K$_2$IrBr$_6$ and 2.2\,eV for K$_2$IrCl$_6$~\cite{khan2019} confirms reduced electronic correlations in the bromide compared to the chloride. This result is cross-checked by analyzing ligand contributions to the density of states near the Fermi level in the uncorrelated band structure. We find 51\% of Br $4p$ states in K$_2$IrBr$_6$, 43\% of Cl $3p$ states in K$_2$IrCl$_6$, and 32\% of O $2p$ states in Ba$_2$CeIrO$_6$ as the best oxide analog of antifluorite-type hexahalides. This indicates the increase in covalency from oxides to chlorides and bromides of Ir$^{4+}$. {\cred Another experimental signature of strong covalency in iridium halides is the Ir$^{4+}$ magnetic form-factor that shows acute deviations from the ionic value in the case of K$_2$IrCl$_6$~\cite{lynn1976}.}

\section{Low-temperature magnetism}
\label{sec:magnetism}

\subsection{Magnetization}

\begin{figure}
	\centering
	\includegraphics[scale=0.5]{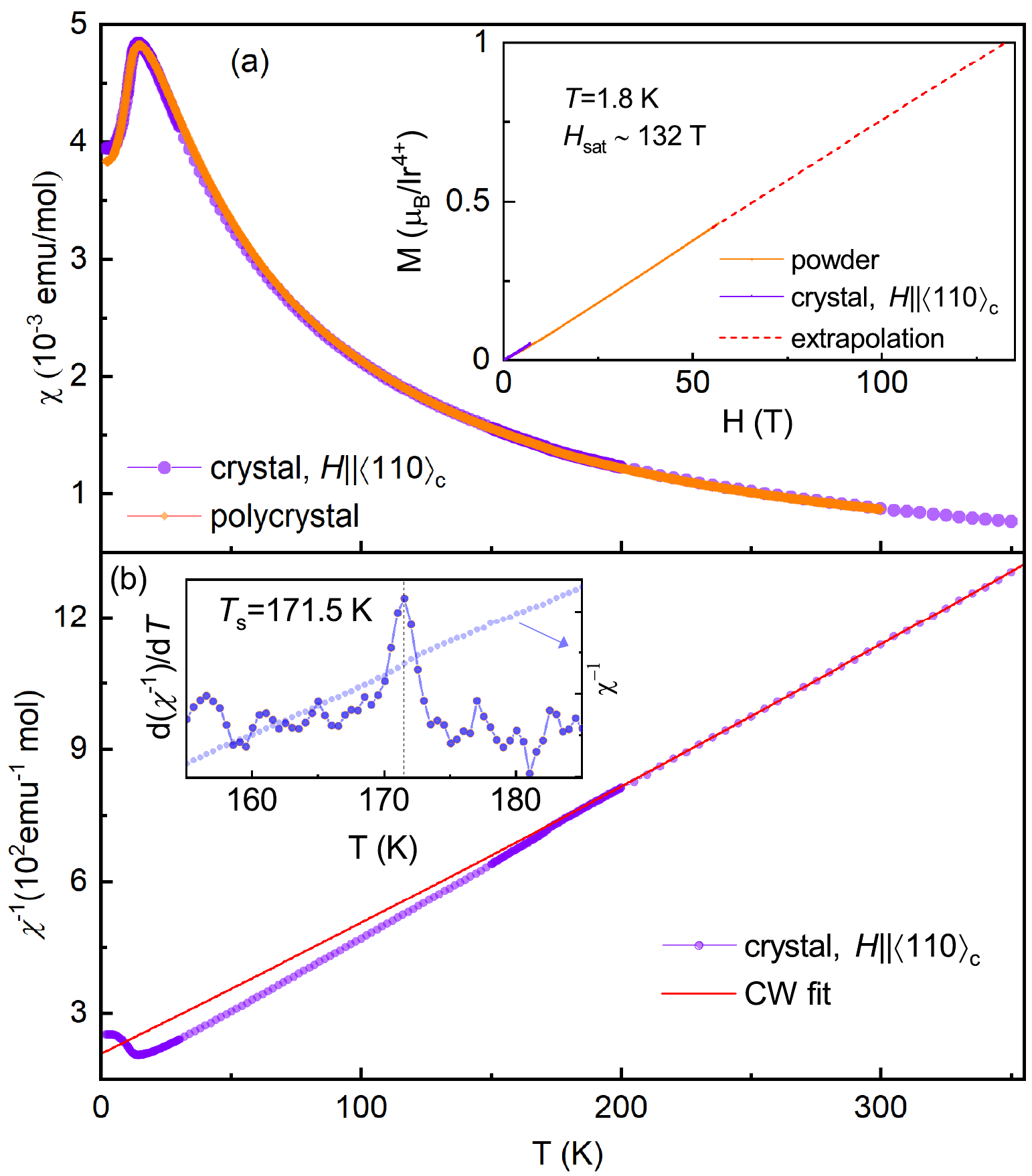}
	\caption{(a) Temperature dependence of dc-magnetic susceptibility ($\chi$) measured on powder ($\mu_0H=0.1$\,T) and on a stack of co-aligned single crystals with the field applied along $\langle110\rangle_c$ ($\mu_0H=3$\,T). The inset shows field dependence of dc-magnetization measured at 1.8\,K in pulsed fields (on powder), and the dotted line is the linear extrapolation to the saturation magnetization of $1\,\mu_{\rm B}$ expected for $j=\frac12$ and $g=2$. (b) $T$-dependence of the inverse susceptibiltiy $\chi ^{-1}$ for the single crystal assembly. The solid red line is the Curie-Weiss (CW) fit to the $\chi^{-1}(T)$ data at 215 $\leq T \leq$ 350 K. The inset shows the $\chi^{-1}(T)$ and d($\chi^{-1}(T)$)/d$T$ in the vicinity of the $\alpha-\beta$ structural phase transition.}\label{fig:chi}
\end{figure}

Magnetic response of K$_2$IrBr$_6$ is nearly isotropic. In Fig.~\ref{fig:chi}a, we show that magnetic susceptibility data collected on powder and on a stack of co-aligned single crystals are nearly indistinguishable. Individual single crystals had the weight of $0.1-0.2$\,mg and were too small for a high-precision measurement up to room temperature. Nevertheless, with the accuracy feasible for such a small sample no directional dependence of the magnetization could be observed either~\cite{suppl}. This result is consistent with the isotropic nature of nearly $j_{\rm eff}=\frac12$ moments of Ir$^{4+}$.

Inverse susceptibility measured on single crystals reveals a small step at the $\alpha-\beta$ transition around 170\,K (Fig.~\ref{fig:chi}b). We used the temperature range above the transition for a Curie-Weiss fit,
\begin{equation}
\chi(T)=\chi_{0} + \frac{N_{\rm A} \mu_{\rm eff}^2}{3k_{\rm B}(T-\Theta_{\rm CW})},\label{Eq.6}
\end{equation}
where $\chi_0=\chi_{\rm dia}+\chi_{\rm vV}$ is the temperature-independent contribution due to core diamagnetism and van Vleck paramagnetism, $N_{\rm A}$ is Avogadro's number, $k_{\rm B}$ is Boltzmann constant, $\mu_{\rm eff}$ is paramagnetic effective moment, and $\Theta_{\rm CW}$ is the Curie-Weiss temperature. The fitted value of $\chi_0=-7.08\times 10^{-5}$\,emu/mol can be compared to $\chi_{\rm dia}=-2.66\times 10^{-4}$\,emu/mol~\cite{bain2008} and yields $\chi_{\rm vV}=1.95\times 10^{-4}$\,emu/mol, which is on par with the van Vleck terms in other Ir$^{4+}$ compounds~\cite{khan2019,majumder2019}. 

The fitted value of paramagnetic effective moment, $\mu_{\rm eff}=1.68$\,$\mu_{\rm B}$, is slightly lower than 1.73\,$\mu_B$ expected for the $j_{\rm eff}=\frac12$ state with $g=2$. The fitting is done in the temperature range of the cubic $\alpha$-phase that features extensive local disorder (Sec.~\ref{sec:structure}), but we deem this disorder unlikely to influence $\mu_{\rm eff}$, because similar dynamic disorder has been also reported for K$_2$IrCl$_6$, where the expected value of $\mu_{\rm eff}=1.73$\,$\mu_B$ is readily obtained from the Curie-Weiss fitting~\cite{khan2019}. Moreover, the electronic state of Ir$^{4+}$ in K$_2$IrBr$_6$ remains very close to $j_{\rm eff}=\frac12$ even in the $\alpha$-phase (Sec.~\ref{sec:rixs}). Another possibility is that the temperature range of the fit was not high enough to reach the Curie-Weiss regime. Indeed, $\Theta_{\rm CW}=-72.7$\,K extracted from the fit is nearly twice higher than $-42.6$\,K reported in K$_2$IrCl$_6$~\cite{khan2019}, and suggests that temperatures well above 170\,K may be required for an accurate Curie-Weiss fitting.

The $\Theta_{\rm CW}$ value gauges the strength of magnetic interactions. Its increase in the bromide compared to the chloride has been cross-checked by high-field magnetization measurements. In the inset of Fig.~\ref{fig:chi}, we show the magnetization measured up to 57\,T in pulsed field and extrapolated to the expected saturation magnetization of 1\,$\mu_B$/f.u. The resulting saturation field of 132\,T is indeed much higher than 87\,T estimated for K$_2$IrCl$_6$~\cite{khan2019}. We conclude that the substitution of Cl by Br increases magnetic interactions by $50-70$\%.

The $\chi(T)$ maximum around 14\,K indicates the onset of antiferromagnetic order. The exact N\'eel temperature $T_N=11.7$\,K was determined from the peak in Fisher's heat capacity $d(\chi T)/dT$ in agreement with the specific heat data (Fig.~\ref{fig:heat}). Below $T_N$, the susceptibility remains nearly isotropic, with no significant difference observed when the field is applied along $\langle 100\rangle_c$, $\langle 110\rangle_c$, and $\langle 111\rangle_c$~\cite{suppl}. This seeming absence of magnetic anisotropy even in the long-range ordered state may be related to the formation of magnetic domains and/or to the possible twinning of the crystal upon the structural phase transitions. 

\subsection{Specific heat}
\label{sec:heat}
\begin{figure}
	\centering
	\includegraphics[scale=0.5]{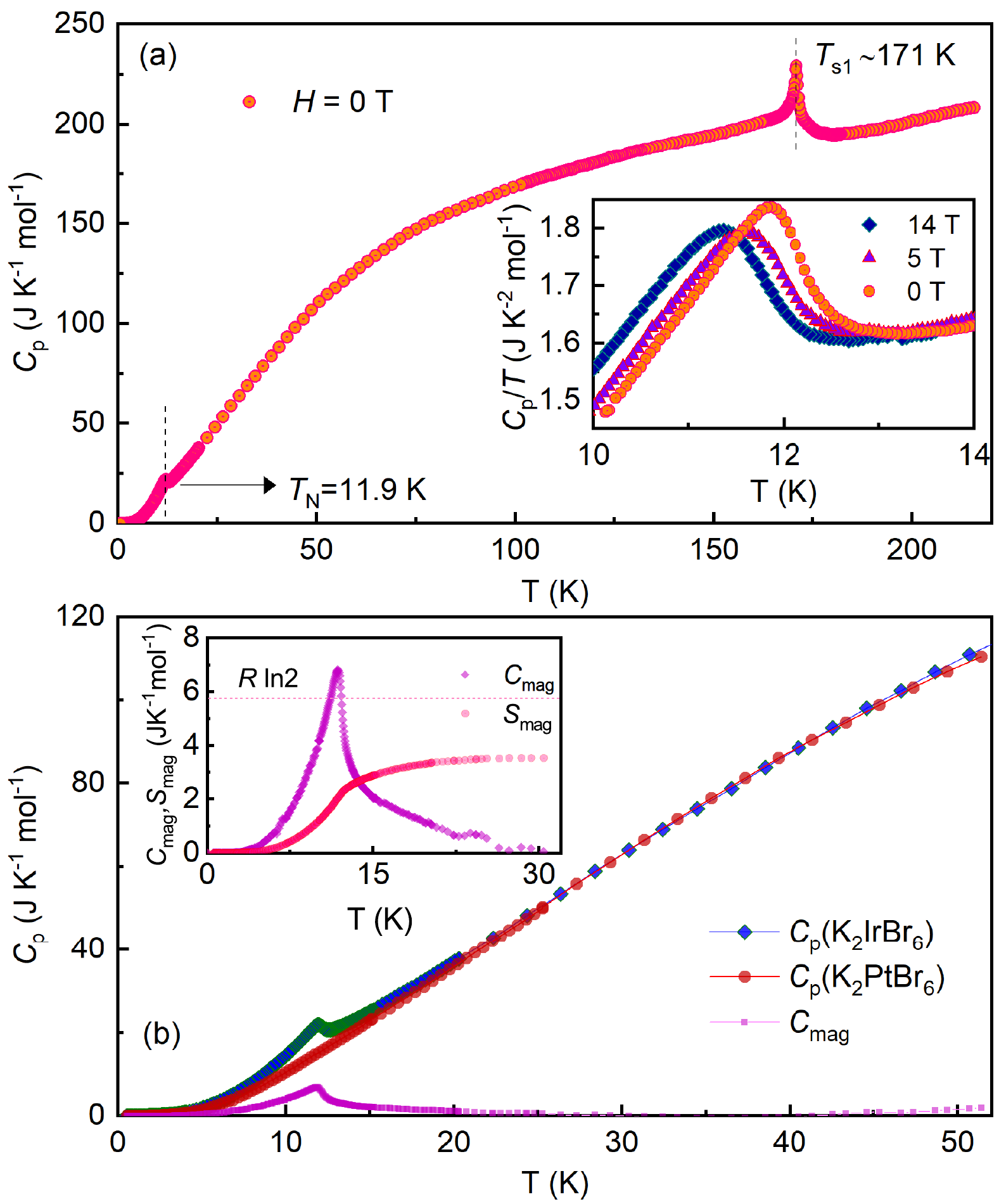}
	\caption{(a) Temperature dependence of zero-field specific heat ($C_p$) measured on a single crystal from 220 K down to 0.5 K exhibits two anomalies at $T_{\rm s1}$ and $T_{\rm N}$. The inset shows $C_{\rm p}$/$T$ as a function of $T$ for $\mu_0H=0$, 5, and 14\, with $H\vert\vert\langle 111 \rangle_c$. (b) Temperature dependence of the specific heat $C_p$ of the non-magnetic K$_2$PtBr$_6$, used as the reference data to subtract the lattice contribution. The evaluated magnetic contribution $C_{\rm mag}$ and the estimated magnetic entropy $S_{\rm mag}$ are shown in the inset for zero field. The dotted line refers to the magnetic entropy of $S_{\rm mag}=R\ln(2j+1)$ for $j=\frac12$.}\label{fig:heat}
\end{figure}

Temperature-dependent specific heat tracks the $\alpha-\beta$ transition around 170\,K and the magnetic ordering transition at 11.9\,K. Symmetric shape of the former peak is consistent with the first-order character of the structural phase transition. On the other hand, the peak at 12\,K is reminiscent of a $\lambda$-type anomaly and suggests second-order nature of the transition, which is confirmed by the absence of hysteresis in thermal expansion~\footnote{Specific heat does not show thermal hysteresis at any of the transitions because of long heat pulses used in the relaxation method.}. We thus expect that no significant structural changes occur in K$_2$IrBr$_6$ upon magnetic ordering. Applied magnetic field shifts the transition toward lower temperatures (inset of Fig.~\ref{fig:heat}a).

Lattice contribution was estimated by measuring specific heat of non-magnetic K$_2$PtBr$_6$. Magnetic specific heat $C_{\rm mag}$ of K$_2$IrBr$_6$ obtained by subtraction was used to estimate magnetic entropy as 
\begin{equation}
S_{\rm mag}(T)=\int_{0}^{T}\frac{C_{\rm mag}}{T'}\,dT'.\label{Eq.7}
\end{equation}   
The resulting magnetic entropy is shown in the inset of Fig.~\ref{fig:heat}b and reaches only 63\,\% of $R\ln 2$ expected for $j_{\rm eff}=\frac12$ moments. We posit that K$_2$PtBr$_6$ may not be the best reference compound for K$_2$IrBr$_6$, because even isostructural compounds of adjacent chemical elements sometimes show non-matching phonon spectra and, therefore, different lattice contributions to the specific heat~\cite{widmann2019}. The presence of soft phonon modes and structural phase transitions exacerbates this situation. Indeed, the Curie-Weiss temperature of $-72.7$\,K suggests that a fraction of magnetic entropy should be released above 30\,K, but this can't be seen when K$_2$PtBr$_6$ data are used as the reference.

We nevertheless expect that our lattice contribution should be rather accurate at low temperatures. Only 50\% of total magnetic entropy is released below $T_N$, indicating that magnetic frustration is not insignificant, similar to some of the Ir$^{4+}$ double perovskites that also feature a distorted version of the frustrated fcc spin lattice~\cite{cao2013}.  

\subsection{Microscopic magnetic model}
\label{sec:model}
Exchange couplings in Ir$^{4+}$ compounds are described by the spin Hamiltonian
\begin{equation}
 \mathcal H=\sum_{\langle ij\rangle} \mathbf S_i\,\mathbb J_{ij}\,\mathbf S_j,
\end{equation}
with $S=\frac12$ and the sum taken over atomic pairs. The $\mathbb J_{ij}$'s are exchange tensors of the form:
\begin{equation}
\mathbb J_{ij}=
\begin{pmatrix}
J+\Gamma_{xx} & D_{z}+\Gamma_{xy}  & -D_{y}+\Gamma_{xz} \\
  -D_{z}+\Gamma_{xy} & J+\Gamma_{yy} & D_{x}+\Gamma_{yz} \\
  D_{y}+\Gamma_{xz} & -D_{x}+\Gamma_{yz} & J+\Gamma_{zz}
\end{pmatrix}
,
\notag\end{equation}
where $J$ is the isotropic (Heisenberg) coupling, $\mathbf D=(D_x,D_y,D_z)$ is the Dzyaloshinskii-Moriya (DM) interaction vector, and $\Gamma$ describes the symmetric portion of anisotropic exchange.

In the cubic structure of K$_2$IrCl$_6$, each exchange tensor is reduced to the simpler form 
\begin{equation}
\mathbb J_{xy}=
\begin{pmatrix}
J & \pm \Gamma  &0 \\
  \pm \Gamma & J & 0\\
  0 & 0 & J+K
\end{pmatrix}
\label{eq:cubic}
\end{equation}
where {\cred diagonal part of the anisotropic exchange is reduced to the isotropic Kitaev term $K$, off-diagonal part is restricted to the single component $\Gamma$, and DM interactions are forbidden by symmetry. The exchange parameters} $J\simeq 13$\,K, $K\simeq 5$\,K, and $\Gamma\simeq 1$\,K were determined from superexchange theory of Refs.~\onlinecite{rau2014,winter2016} using the optimal value of $U=2.2$\,eV. 

Full exchange tensors determined for the 20\,K crystal structure of $\gamma$-K$_2$IrBr$_6$ using the same method and $U=1.8$\,eV are given in the Supplemental Material~\cite{suppl}. {\cred In both cases, hopping parameters were obtained by Wannier projections that take covalency effects into account.} Lower symmetry of K$_2$IrBr$_6$ causes exchange tensors to deviate from the simple form of Eq.~\eqref{eq:cubic}, although one can still recognize a similar structure and introduce effective values of {\cred $J$ as the isotropic coupling, $K$ as the leading diagonal anisotropy, and $\Gamma$ as the leading symmetric off-diagonal anisotropy,} augmented by Dzyaloshinskii-Moriya vectors $\Dv$ for bonds $\mathbb J_5-\mathbb J_8$ without inversion symmetry.

\begin{table}
\caption{\label{tab:exchange}
The Ir--Ir distances (in\,\r A) and main components of the exchange tensors (in\,K) calculated for the 20\,K structure of $\gamma$-K$_2$IrBr$_6$. $|\Dv|$ is the length of the Dzyaloshinskii-Moriya vector. For the notation of individual interactions see Fig.~\ref{fig:spin}.
}
\begin{ruledtabular}
\begin{tabular}{c@{\hspace{0.6cm}}c@{\hspace{0.6cm}}ccrc}
 & $d_{\rm Ir-Ir}$ & $J$ & $K$ & $\Gamma$ & $|\Dv|$ \\\hline
 $\mathbb J_1$ & 7.132 & 20.0 & 8.9 & 1.5 & 0\\
 $\mathbb J_2$ & 7.131 & 15.8 & 7.4 & $-0.9$ & 0 \\
 $\mathbb J_3$ & 7.132 & 20.0 & 8.9 & 1.5 & 0 \\
 $\mathbb J_4$ & 7.131 & 15.8 & 7.4 & $-0.9$ & 0 \\
 $\mathbb J_5$ & 7.243 & 15.7 & 5.7 & 2.1 & 12.1 \\
 $\mathbb J_6$ & 7.269 & 14.0 & 4.6 & $-3.0$ & 13.3 \\
 $\mathbb J_7$ & 7.269 & 14.0 & 4.6 & 3.0 & 13.8 \\
 $\mathbb J_8$ & 7.243 & 15.7 & 5.7 & $-2.1$ & 12.1 \\
\end{tabular}
\end{ruledtabular}
\end{table}

\begin{figure}
\includegraphics[scale=0.98]{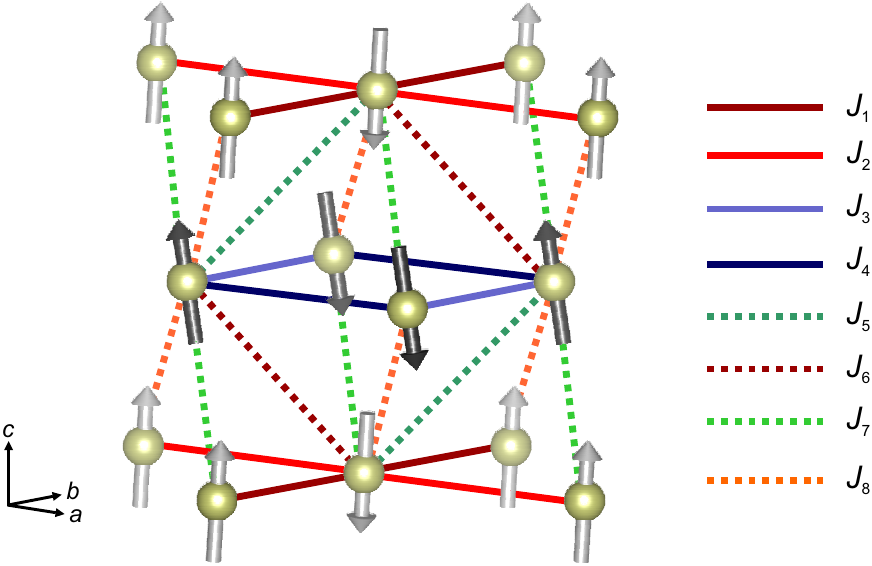}
\caption{\label{fig:spin}
Distorted fcc spin lattice in the $\gamma$-phase of K$_2$IrBr$_6$. Crystallographic directions of the monoclinic $\gamma$-structure are shown. The spin configuration of canted type-I order, as obtained from DFT+$U$+SO calculations, is superimposed (see text for details).
}
\end{figure}

The monoclinic structure of $\gamma$-K$_2$IrBr$_6$ features 8 nonequivalent nearest-neighbor exchange pathways shown in Fig.~\ref{fig:spin}. The corresponding interactions are listed in Table~\ref{tab:exchange}. In contrast to the regular fcc lattice, each triangular loop includes three couplings of different strength. This partially alleviates the frustration and explains the change in the frustration ratio $\Theta_{\rm CW}/T_N$ from 13.7 in K$_2$IrCl$_6$~\cite{khan2019} to 6.0 in K$_2$IrBr$_6$. 

We also find up to 50\% increase in the magnitude of exchange couplings compared to K$_2$IrCl$_6$. This increase should be traced back to the reduced strength of electronic correlations and, eventually, to the enhanced {\cred covalency of the Ir--Br interactions (see Sec.~\ref{sec:correlations})}. The hierarchy of hopping parameters does not change significantly. Therefore, all interactions $\mathbb J_1-\mathbb J_8$ feature large Heisenberg exchange with a weaker antiferromagnetic Kitaev exchange and minor off-diagonal anisotropy. 

{\cred The fcc antiferromagnets with predominant nearest-neighbor couplings feature two types of collinear magnetic order known as type-I and type-III~\cite{haar1962}. 
We computed energies of type-I and type-III spin configurations of $\gamma$-K$_2$IrBr$_6$ in DFT+$U$+SO (in \texttt{VASP}) while allowing magnetic moment directions to relax. This led to the 1.6\,meV/f.u. lower energy of the \mbox{type-I} order (Fig.~\ref{fig:spin}), in agreement with the neutron results that revealed the propagation vector $\mathbf k=(\frac12,\frac12,0)$ compatible with the type-I order. Stabilization of this order is accompanied by a spin canting wherein four sublattices feature spins directed along $\pm(-0.075,-0.050,0.996)$ and $\pm(0.254,-0.085,0.963)$ within the Cartesian coordinate frame of $\gamma$-K$_2$IrBr$_6$. The sizable canting angle of $13.7^{\circ}$ is probably a result of large DM couplings on some of the bonds.
}


\section{Discussion and Summary}
\label{sec:discussion}
The non-cubic crystal-field splitting of $\Delta=50-70$\,meV observed in K$_2$IrBr$_6$ {\cred and other Ir$^{4+}$ antifluorites~\cite{plessis2020}} is remarkably smaller than in Ir$^{4+}$ oxides with weakly distorted IrO$_6$ octahedra. Larger deformations of the octahedra enhance $\Delta$ significantly. For example, post-perovskite CaIrO$_3$ reveals $\Delta=-0.71$\,eV~\cite{moretti2014} for the tetragonal distortion of $\delta_t=-1.9$\%. Our calculations (Fig.~\ref{fig:delta-cal}) suggest that a similar distortion in the bromide would lead to only a small $\Delta\simeq 0.01$\,eV. 

This much weaker crystal-field splitting can be traced back to the reduced charge of the ligand and the enhanced covalency that both weaken electrostatic effects on the Ir $5d$ orbitals. The crystal-field splitting as low as 50\,meV can be reached in this way. Its size appears to be determined not only by deformations of the IrBr$_6$ octahedron, but also by the positions of more distant neighbors such as K$^+$ ions. On one hand, this creates a large flexibility. Different types of distortions compensate each other and keep $\Delta$ small, as in the $\gamma$-phase of K$_2$IrBr$_6$. On the other hand, residual crystal-field splittings are hard to eliminate in this setting, because virtually every structural distortion, be it static in monoclinic $\gamma$-K$_2$IrBr$_6$ or dynamic in cubic $\alpha$-K$_2$IrBr$_6$, renders $\Delta$ non-zero. Reaching $\Delta=0$ and the pure, unperturbed $j_{\rm eff}=\frac12$ state appears to be exceedingly difficult, because even effects like zero-point motions and vibronic couplings are likely to affect crystal-field levels of Ir$^{4+}$.

In summary, we reported and elucidated structural transformations in K$_2$IrBr$_6$ and their effect on the non-cubic crystal-field splitting $\Delta$ for the Ir$^{4+}$ ion. Relevant structural distortions are separated into two groups. Tensile tetragonal strain and octahedral rotations on one side and octahedral tilts as well as orthorhombic deformations of the IrBr$_6$ octahedra on the other side lead to antagonistic changes in $\Delta$. These changes underlie the non-monotonic evolution of $\Delta$ across the structural phase transitions. The transitions cause a distortion of the fcc spin lattice and partially alleviate magnetic frustration. Magnetic couplings are similar in nature to the cubic Ir$^{4+}$ hexahalides, with the predominant Heisenberg term and a weak antiferromagnetic Kitaev term, but show a larger magnitude thanks to the reduced correlations on the Ir site. This way, covalency not only reduces crystal-field splittings, but also enhances magnetic couplings. {\cred \mbox{Type-I} magnetic order is stabilized, albeit with a large spin canting caused by the underlying structural distortion.}


\acknowledgments
NK thanks Somnath Ghara for his help with resistivity measurements. AAT thanks Adam Aczel and Anna Efimenko for fruitful discussions on the Ir$^{4+}$ antifluorites, and Yurii Skourski for performing the high-field magnetization measurements. The work in Augsburg was supported by the Federal Ministry for Education and Research through the Sofja Kovalevskaya Award of Alexander von Humboldt Foundation (AAT). The work was partially supported by the Ministry of Science and Higher Education of the Russian Federation (through the basic part of the government mandate, Project No. FEUZ-2020-0060). 

We acknowledge ESRF and APS for providing synchrotron beamtime for this project, and thank Andy Fitch for his technical support during the experiment at ID22, ESRF. This research used resources of the Advanced Photon Source, a U.S. Department of Energy (DOE) Office of Science User Facility, operated for the DOE Office of Science by Argonne National Laboratory under Contract No. DE-AC02-06CH11357. Extraordinary facility operations were supported in part by the DOE Office of Science through the National Virtual Biotechnology Laboratory, a consortium of DOE national laboratories focused on the response to COVID-19, with funding provided by the Coronavirus CARES Act. We also acknowledge the support of the HLD at HZDR, member of European Magnetic Field Laboratory (EMFL).


\begin{thebibliography}{52}%
\makeatletter
\providecommand \@ifxundefined [1]{%
 \@ifx{#1\undefined}
}%
\providecommand \@ifnum [1]{%
 \ifnum #1\expandafter \@firstoftwo
 \else \expandafter \@secondoftwo
 \fi
}%
\providecommand \@ifx [1]{%
 \ifx #1\expandafter \@firstoftwo
 \else \expandafter \@secondoftwo
 \fi
}%
\providecommand \natexlab [1]{#1}%
\providecommand \enquote  [1]{``#1''}%
\providecommand \bibnamefont  [1]{#1}%
\providecommand \bibfnamefont [1]{#1}%
\providecommand \citenamefont [1]{#1}%
\providecommand \href@noop [0]{\@secondoftwo}%
\providecommand \href [0]{\begingroup \@sanitize@url \@href}%
\providecommand \@href[1]{\@@startlink{#1}\@@href}%
\providecommand \@@href[1]{\endgroup#1\@@endlink}%
\providecommand \@sanitize@url [0]{\catcode `\\12\catcode `\$12\catcode
  `\&12\catcode `\#12\catcode `\^12\catcode `\_12\catcode `\%12\relax}%
\providecommand \@@startlink[1]{}%
\providecommand \@@endlink[0]{}%
\providecommand \url  [0]{\begingroup\@sanitize@url \@url }%
\providecommand \@url [1]{\endgroup\@href {#1}{\urlprefix }}%
\providecommand \urlprefix  [0]{URL }%
\providecommand \Eprint [0]{\href }%
\providecommand \doibase [0]{https://doi.org/}%
\providecommand \selectlanguage [0]{\@gobble}%
\providecommand \bibinfo  [0]{\@secondoftwo}%
\providecommand \bibfield  [0]{\@secondoftwo}%
\providecommand \translation [1]{[#1]}%
\providecommand \BibitemOpen [0]{}%
\providecommand \bibitemStop [0]{}%
\providecommand \bibitemNoStop [0]{.\EOS\space}%
\providecommand \EOS [0]{\spacefactor3000\relax}%
\providecommand \BibitemShut  [1]{\csname bibitem#1\endcsname}%
\let\auto@bib@innerbib\@empty
\bibitem [{\citenamefont {Witczak-Krempa}\ \emph {et~al.}(2014)\citenamefont
  {Witczak-Krempa}, \citenamefont {Chen}, \citenamefont {Kim},\ and\
  \citenamefont {Balents}}]{krempa2014}%
  \BibitemOpen
  \bibfield  {author} {\bibinfo {author} {\bibfnamefont {W.}~\bibnamefont
  {Witczak-Krempa}}, \bibinfo {author} {\bibfnamefont {G.}~\bibnamefont
  {Chen}}, \bibinfo {author} {\bibfnamefont {Y.~B.}\ \bibnamefont {Kim}},\ and\
  \bibinfo {author} {\bibfnamefont {L.}~\bibnamefont {Balents}},\ }\bibfield
  {title} {\bibinfo {title} {Correlated quantum phenomena in the strong
  spin-orbit regime},\ }\href
  {https://doi.org/10.1146/annurev-conmatphys-020911-125138} {\bibfield
  {journal} {\bibinfo  {journal} {Ann. Rev. Condens. Matter Phys.}\ }\textbf
  {\bibinfo {volume} {5}},\ \bibinfo {pages} {57} (\bibinfo {year}
  {2014})}\BibitemShut {NoStop}%
\bibitem [{\citenamefont {Rau}\ \emph {et~al.}(2016)\citenamefont {Rau},
  \citenamefont {Lee},\ and\ \citenamefont {Kee}}]{rau2016}%
  \BibitemOpen
  \bibfield  {author} {\bibinfo {author} {\bibfnamefont {J.~G.}\ \bibnamefont
  {Rau}}, \bibinfo {author} {\bibfnamefont {E.~K.-H.}\ \bibnamefont {Lee}},\
  and\ \bibinfo {author} {\bibfnamefont {H.-Y.}\ \bibnamefont {Kee}},\
  }\bibfield  {title} {\bibinfo {title} {Spin-orbit physics giving rise to
  novel phases in correlated systems: Iridates and related materials},\ }\href
  {https://doi.org/10.1146/annurev-conmatphys-031115-011319} {\bibfield
  {journal} {\bibinfo  {journal} {Ann. Rev. Condens. Matter Phys.}\ }\textbf
  {\bibinfo {volume} {7}},\ \bibinfo {pages} {195} (\bibinfo {year}
  {2016})}\BibitemShut {NoStop}%
\bibitem [{\citenamefont {Kanungo}\ \emph {et~al.}(2015)\citenamefont
  {Kanungo}, \citenamefont {Yan}, \citenamefont {Merz}, \citenamefont
  {Felser},\ and\ \citenamefont {Jansen}}]{kanungo2015}%
  \BibitemOpen
  \bibfield  {author} {\bibinfo {author} {\bibfnamefont {S.}~\bibnamefont
  {Kanungo}}, \bibinfo {author} {\bibfnamefont {B.}~\bibnamefont {Yan}},
  \bibinfo {author} {\bibfnamefont {P.}~\bibnamefont {Merz}}, \bibinfo {author}
  {\bibfnamefont {C.}~\bibnamefont {Felser}},\ and\ \bibinfo {author}
  {\bibfnamefont {M.}~\bibnamefont {Jansen}},\ }\bibfield  {title} {\bibinfo
  {title} {{Na$_4$IrO$_4$}: square-planar coordination of a transition metal in
  $d^5$ configuration due to weak on-site {Coulomb} interactions},\ }\href
  {https://doi.org/10.1002/anie.201411959} {\bibfield  {journal} {\bibinfo
  {journal} {Angew. Chem. Int. Ed.}\ }\textbf {\bibinfo {volume} {54}},\
  \bibinfo {pages} {5417} (\bibinfo {year} {2015})}\BibitemShut {NoStop}%
\bibitem [{\citenamefont {Ming}\ \emph {et~al.}(2017)\citenamefont {Ming},
  \citenamefont {Autieri}, \citenamefont {Yamauchi},\ and\ \citenamefont
  {Picozzi}}]{ming2017}%
  \BibitemOpen
  \bibfield  {author} {\bibinfo {author} {\bibfnamefont {X.}~\bibnamefont
  {Ming}}, \bibinfo {author} {\bibfnamefont {C.}~\bibnamefont {Autieri}},
  \bibinfo {author} {\bibfnamefont {K.}~\bibnamefont {Yamauchi}},\ and\
  \bibinfo {author} {\bibfnamefont {S.}~\bibnamefont {Picozzi}},\ }\bibfield
  {title} {\bibinfo {title} {Role of square planar coordination in the magnetic
  properties of {Na$_4$IrO$_4$}},\ }\href
  {https://doi.org/10.1103/PhysRevB.96.205158} {\bibfield  {journal} {\bibinfo
  {journal} {Phys. Rev. B}\ }\textbf {\bibinfo {volume} {96}},\ \bibinfo
  {pages} {205158} (\bibinfo {year} {2017})}\BibitemShut {NoStop}%
\bibitem [{\citenamefont {Winter}\ \emph {et~al.}(2017)\citenamefont {Winter},
  \citenamefont {Tsirlin}, \citenamefont {Daghofer}, \citenamefont {{van den
  Brink}}, \citenamefont {Singh}, \citenamefont {Gegenwart},\ and\
  \citenamefont {Valent{\'\i}}}]{winter2017}%
  \BibitemOpen
  \bibfield  {author} {\bibinfo {author} {\bibfnamefont {S.~M.}\ \bibnamefont
  {Winter}}, \bibinfo {author} {\bibfnamefont {A.~A.}\ \bibnamefont {Tsirlin}},
  \bibinfo {author} {\bibfnamefont {M.}~\bibnamefont {Daghofer}}, \bibinfo
  {author} {\bibfnamefont {J.}~\bibnamefont {{van den Brink}}}, \bibinfo
  {author} {\bibfnamefont {Y.}~\bibnamefont {Singh}}, \bibinfo {author}
  {\bibfnamefont {P.}~\bibnamefont {Gegenwart}},\ and\ \bibinfo {author}
  {\bibfnamefont {R.}~\bibnamefont {Valent{\'\i}}},\ }\bibfield  {title}
  {\bibinfo {title} {Models and materials for generalized {Kitaev} magnetism},\
  }\href {https://doi.org/10.1088/1361-648X/aa8cf5} {\bibfield  {journal}
  {\bibinfo  {journal} {J. Phys.: Condens. Matter}\ }\textbf {\bibinfo {volume}
  {29}},\ \bibinfo {pages} {493002} (\bibinfo {year} {2017})}\BibitemShut
  {NoStop}%
\bibitem [{\citenamefont {Cao}\ and\ \citenamefont
  {Schlottmann}(2018)}]{cao2018}%
  \BibitemOpen
  \bibfield  {author} {\bibinfo {author} {\bibfnamefont {G.}~\bibnamefont
  {Cao}}\ and\ \bibinfo {author} {\bibfnamefont {P.}~\bibnamefont
  {Schlottmann}},\ }\bibfield  {title} {\bibinfo {title} {The challenge of
  spin-orbit-tuned ground states in iridates: a key issues review},\ }\href
  {https://doi.org/10.1088/1361-6633/aaa979} {\bibfield  {journal} {\bibinfo
  {journal} {Rep. Prog. Phys.}\ }\textbf {\bibinfo {volume} {81}},\ \bibinfo
  {pages} {042502} (\bibinfo {year} {2018})}\BibitemShut {NoStop}%
\bibitem [{\citenamefont {Bertinshaw}\ \emph {et~al.}(2019)\citenamefont
  {Bertinshaw}, \citenamefont {Kim}, \citenamefont {Khaliullin},\ and\
  \citenamefont {Kim}}]{bertinshaw2019}%
  \BibitemOpen
  \bibfield  {author} {\bibinfo {author} {\bibfnamefont {J.}~\bibnamefont
  {Bertinshaw}}, \bibinfo {author} {\bibfnamefont {Y.~K.}\ \bibnamefont {Kim}},
  \bibinfo {author} {\bibfnamefont {G.}~\bibnamefont {Khaliullin}},\ and\
  \bibinfo {author} {\bibfnamefont {B.~J.}\ \bibnamefont {Kim}},\ }\bibfield
  {title} {\bibinfo {title} {Square lattice iridates},\ }\href
  {https://doi.org/10.1146/annurev-conmatphys-031218-013113} {\bibfield
  {journal} {\bibinfo  {journal} {Ann. Rev. Condens. Matter Phys.}\ }\textbf
  {\bibinfo {volume} {10}},\ \bibinfo {pages} {315} (\bibinfo {year}
  {2019})}\BibitemShut {NoStop}%
\bibitem [{\citenamefont {Gretarsson}\ \emph {et~al.}(2013)\citenamefont
  {Gretarsson}, \citenamefont {Clancy}, \citenamefont {Liu}, \citenamefont
  {Hill}, \citenamefont {Bozin}, \citenamefont {Singh}, \citenamefont {Manni},
  \citenamefont {Gegenwart}, \citenamefont {Kim}, \citenamefont {Said},
  \citenamefont {Casa}, \citenamefont {Gog}, \citenamefont {Upton},
  \citenamefont {Kim}, \citenamefont {Yu}, \citenamefont {Katukuri},
  \citenamefont {Hozoi}, \citenamefont {{van den Brink}},\ and\ \citenamefont
  {Kim}}]{gretarsson2013}%
  \BibitemOpen
  \bibfield  {author} {\bibinfo {author} {\bibfnamefont {H.}~\bibnamefont
  {Gretarsson}}, \bibinfo {author} {\bibfnamefont {J.~P.}\ \bibnamefont
  {Clancy}}, \bibinfo {author} {\bibfnamefont {X.}~\bibnamefont {Liu}},
  \bibinfo {author} {\bibfnamefont {J.~P.}\ \bibnamefont {Hill}}, \bibinfo
  {author} {\bibfnamefont {E.}~\bibnamefont {Bozin}}, \bibinfo {author}
  {\bibfnamefont {Y.}~\bibnamefont {Singh}}, \bibinfo {author} {\bibfnamefont
  {S.}~\bibnamefont {Manni}}, \bibinfo {author} {\bibfnamefont
  {P.}~\bibnamefont {Gegenwart}}, \bibinfo {author} {\bibfnamefont
  {J.}~\bibnamefont {Kim}}, \bibinfo {author} {\bibfnamefont {A.~H.}\
  \bibnamefont {Said}}, \bibinfo {author} {\bibfnamefont {D.}~\bibnamefont
  {Casa}}, \bibinfo {author} {\bibfnamefont {T.}~\bibnamefont {Gog}}, \bibinfo
  {author} {\bibfnamefont {M.~H.}\ \bibnamefont {Upton}}, \bibinfo {author}
  {\bibfnamefont {H.-S.}\ \bibnamefont {Kim}}, \bibinfo {author} {\bibfnamefont
  {J.}~\bibnamefont {Yu}}, \bibinfo {author} {\bibfnamefont {V.~M.}\
  \bibnamefont {Katukuri}}, \bibinfo {author} {\bibfnamefont {L.}~\bibnamefont
  {Hozoi}}, \bibinfo {author} {\bibfnamefont {J.}~\bibnamefont {{van den
  Brink}}},\ and\ \bibinfo {author} {\bibfnamefont {Y.-J.}\ \bibnamefont
  {Kim}},\ }\bibfield  {title} {\bibinfo {title} {Crystal-field splitting and
  correlation effect on the electronic structure of {A$_2$IrO$_3$}},\ }\href
  {https://doi.org/10.1103/PhysRevLett.110.076402} {\bibfield  {journal}
  {\bibinfo  {journal} {Phys. Rev. Lett.}\ }\textbf {\bibinfo {volume} {110}},\
  \bibinfo {pages} {076402} (\bibinfo {year} {2013})}\BibitemShut {NoStop}%
\bibitem [{\citenamefont {Aczel}\ \emph {et~al.}(2019)\citenamefont {Aczel},
  \citenamefont {Clancy}, \citenamefont {Chen}, \citenamefont {Zhou},
  \citenamefont {{Reig-i-Plessis}}, \citenamefont {MacDougall}, \citenamefont
  {Ruff}, \citenamefont {Upton}, \citenamefont {Islam}, \citenamefont
  {Williams}, \citenamefont {Calder},\ and\ \citenamefont {Yan}}]{aczel2019}%
  \BibitemOpen
  \bibfield  {author} {\bibinfo {author} {\bibfnamefont {A.~A.}\ \bibnamefont
  {Aczel}}, \bibinfo {author} {\bibfnamefont {J.~P.}\ \bibnamefont {Clancy}},
  \bibinfo {author} {\bibfnamefont {Q.}~\bibnamefont {Chen}}, \bibinfo {author}
  {\bibfnamefont {H.~D.}\ \bibnamefont {Zhou}}, \bibinfo {author}
  {\bibfnamefont {D.}~\bibnamefont {{Reig-i-Plessis}}}, \bibinfo {author}
  {\bibfnamefont {G.~J.}\ \bibnamefont {MacDougall}}, \bibinfo {author}
  {\bibfnamefont {J.~P.~C.}\ \bibnamefont {Ruff}}, \bibinfo {author}
  {\bibfnamefont {M.~H.}\ \bibnamefont {Upton}}, \bibinfo {author}
  {\bibfnamefont {Z.}~\bibnamefont {Islam}}, \bibinfo {author} {\bibfnamefont
  {T.~J.}\ \bibnamefont {Williams}}, \bibinfo {author} {\bibfnamefont
  {S.}~\bibnamefont {Calder}},\ and\ \bibinfo {author} {\bibfnamefont {J.-Q.}\
  \bibnamefont {Yan}},\ }\bibfield  {title} {\bibinfo {title} {Revisiting the
  {Kitaev} material candidacy of {Ir$^{4+}$} double perovskite iridates},\
  }\href {https://doi.org/10.1103/PhysRevB.99.134417} {\bibfield  {journal}
  {\bibinfo  {journal} {Phys. Rev. B}\ }\textbf {\bibinfo {volume} {99}},\
  \bibinfo {pages} {134417} (\bibinfo {year} {2019})}\BibitemShut {NoStop}%
\bibitem [{\citenamefont {Revelli}\ \emph {et~al.}(2019)\citenamefont
  {Revelli}, \citenamefont {Loo}, \citenamefont {Kiese}, \citenamefont
  {Becker}, \citenamefont {Fr\"ohlich}, \citenamefont {Lorenz}, \citenamefont
  {{Moretti Sala}}, \citenamefont {Monaco}, \citenamefont {Buessen},
  \citenamefont {Attig}, \citenamefont {Hermanns}, \citenamefont {Streltsov},
  \citenamefont {Khomskii}, \citenamefont {{van den Brink}}, \citenamefont
  {Braden}, \citenamefont {{van Loosdrecht}}, \citenamefont {Trebst},
  \citenamefont {Paramekanti},\ and\ \citenamefont
  {Gr\"uninger}}]{revelli2019}%
  \BibitemOpen
  \bibfield  {author} {\bibinfo {author} {\bibfnamefont {A.}~\bibnamefont
  {Revelli}}, \bibinfo {author} {\bibfnamefont {C.~C.}\ \bibnamefont {Loo}},
  \bibinfo {author} {\bibfnamefont {D.}~\bibnamefont {Kiese}}, \bibinfo
  {author} {\bibfnamefont {P.}~\bibnamefont {Becker}}, \bibinfo {author}
  {\bibfnamefont {T.}~\bibnamefont {Fr\"ohlich}}, \bibinfo {author}
  {\bibfnamefont {T.}~\bibnamefont {Lorenz}}, \bibinfo {author} {\bibfnamefont
  {M.}~\bibnamefont {{Moretti Sala}}}, \bibinfo {author} {\bibfnamefont
  {G.}~\bibnamefont {Monaco}}, \bibinfo {author} {\bibfnamefont {F.~L.}\
  \bibnamefont {Buessen}}, \bibinfo {author} {\bibfnamefont {J.}~\bibnamefont
  {Attig}}, \bibinfo {author} {\bibfnamefont {M.}~\bibnamefont {Hermanns}},
  \bibinfo {author} {\bibfnamefont {S.~V.}\ \bibnamefont {Streltsov}}, \bibinfo
  {author} {\bibfnamefont {D.~I.}\ \bibnamefont {Khomskii}}, \bibinfo {author}
  {\bibfnamefont {J.}~\bibnamefont {{van den Brink}}}, \bibinfo {author}
  {\bibfnamefont {M.}~\bibnamefont {Braden}}, \bibinfo {author} {\bibfnamefont
  {P.~H.~M.}\ \bibnamefont {{van Loosdrecht}}}, \bibinfo {author}
  {\bibfnamefont {S.}~\bibnamefont {Trebst}}, \bibinfo {author} {\bibfnamefont
  {A.}~\bibnamefont {Paramekanti}},\ and\ \bibinfo {author} {\bibfnamefont
  {M.}~\bibnamefont {Gr\"uninger}},\ }\bibfield  {title} {\bibinfo {title}
  {Spin-orbit entangled $j=\frac12$ moments in {Ba$_2$CeIrO$_6$}: A frustrated
  fcc quantum magnet},\ }\href {https://doi.org/10.1103/PhysRevB.100.085139}
  {\bibfield  {journal} {\bibinfo  {journal} {Phys. Rev. B}\ }\textbf {\bibinfo
  {volume} {100}},\ \bibinfo {pages} {085139} (\bibinfo {year}
  {2019})}\BibitemShut {NoStop}%
\bibitem [{\citenamefont {Rossi}\ \emph {et~al.}(2017)\citenamefont {Rossi},
  \citenamefont {Retegan}, \citenamefont {Giacobbe}, \citenamefont {Fumagalli},
  \citenamefont {Efimenko}, \citenamefont {Kulka}, \citenamefont {Wohlfeld},
  \citenamefont {Gubanov},\ and\ \citenamefont {{Moretti Sala}}}]{rossi2017}%
  \BibitemOpen
  \bibfield  {author} {\bibinfo {author} {\bibfnamefont {M.}~\bibnamefont
  {Rossi}}, \bibinfo {author} {\bibfnamefont {M.}~\bibnamefont {Retegan}},
  \bibinfo {author} {\bibfnamefont {C.}~\bibnamefont {Giacobbe}}, \bibinfo
  {author} {\bibfnamefont {R.}~\bibnamefont {Fumagalli}}, \bibinfo {author}
  {\bibfnamefont {A.}~\bibnamefont {Efimenko}}, \bibinfo {author}
  {\bibfnamefont {T.}~\bibnamefont {Kulka}}, \bibinfo {author} {\bibfnamefont
  {K.}~\bibnamefont {Wohlfeld}}, \bibinfo {author} {\bibfnamefont {A.~I.}\
  \bibnamefont {Gubanov}},\ and\ \bibinfo {author} {\bibfnamefont
  {M.}~\bibnamefont {{Moretti Sala}}},\ }\bibfield  {title} {\bibinfo {title}
  {Possibility to realize spin-orbit-induced correlated physics in iridium
  fluorides},\ }\href {https://doi.org/10.1103/PhysRevB.95.235161} {\bibfield
  {journal} {\bibinfo  {journal} {Phys. Rev. B}\ }\textbf {\bibinfo {volume}
  {95}},\ \bibinfo {pages} {235161} (\bibinfo {year} {2017})}\BibitemShut
  {NoStop}%
\bibitem [{Note1()}]{Note1}%
  \BibitemOpen
  \bibinfo {note} {Note that Ref.~\cite {aczel2019} refers to $\Delta $ as the
  splitting between the RIXS peaks, $\Delta _{\protect \rm exp}$. In contrast,
  our $\Delta $ is the actual splitting between the $t_{2g}$ levels, and
  $\Delta =\protect \frac 32\Delta _{\protect \rm exp}$ in the limit of $\Delta
  \ll \lambda $, where $\lambda \simeq 0.4-0.5$\protect \tmspace +\thinmuskip
  {.1667em}eV is the typical spin-orbit coupling in Ir$^{4+}$
  compounds.}\BibitemShut {Stop}%
\bibitem [{\citenamefont {Lines}(1963)}]{lines1963}%
  \BibitemOpen
  \bibfield  {author} {\bibinfo {author} {\bibfnamefont {M.~E.}\ \bibnamefont
  {Lines}},\ }\bibfield  {title} {\bibinfo {title} {Antiferromagnetism in the
  face-centred cubic lattice by a spin-wave method},\ }\href
  {https://doi.org/10.1098/rspa.1963.0007} {\bibfield  {journal} {\bibinfo
  {journal} {Proc. Royal Soc. A}\ }\textbf {\bibinfo {volume} {271}},\ \bibinfo
  {pages} {105} (\bibinfo {year} {1963})}\BibitemShut {NoStop}%
\bibitem [{\citenamefont {Khan}\ \emph {et~al.}(2019)\citenamefont {Khan},
  \citenamefont {Prishchenko}, \citenamefont {Skourski}, \citenamefont
  {Mazurenko},\ and\ \citenamefont {Tsirlin}}]{khan2019}%
  \BibitemOpen
  \bibfield  {author} {\bibinfo {author} {\bibfnamefont {N.}~\bibnamefont
  {Khan}}, \bibinfo {author} {\bibfnamefont {D.}~\bibnamefont {Prishchenko}},
  \bibinfo {author} {\bibfnamefont {Y.}~\bibnamefont {Skourski}}, \bibinfo
  {author} {\bibfnamefont {V.~G.}\ \bibnamefont {Mazurenko}},\ and\ \bibinfo
  {author} {\bibfnamefont {A.~A.}\ \bibnamefont {Tsirlin}},\ }\bibfield
  {title} {\bibinfo {title} {Cubic symmetry and magnetic frustration on the fcc
  spin lattice in {K$_2$IrCl$_6$}},\ }\href
  {https://doi.org/10.1103/PhysRevB.99.144425} {\bibfield  {journal} {\bibinfo
  {journal} {Phys. Rev. B}\ }\textbf {\bibinfo {volume} {99}},\ \bibinfo
  {pages} {144425} (\bibinfo {year} {2019})}\BibitemShut {NoStop}%
\bibitem [{\citenamefont {Schick}\ \emph {et~al.}(2020)\citenamefont {Schick},
  \citenamefont {Ziman},\ and\ \citenamefont {Zhitomirsky}}]{schick2020}%
  \BibitemOpen
  \bibfield  {author} {\bibinfo {author} {\bibfnamefont {R.}~\bibnamefont
  {Schick}}, \bibinfo {author} {\bibfnamefont {T.}~\bibnamefont {Ziman}},\ and\
  \bibinfo {author} {\bibfnamefont {M.~E.}\ \bibnamefont {Zhitomirsky}},\
  }\bibfield  {title} {\bibinfo {title} {Quantum versus thermal fluctuations in
  the fcc antiferromagnet: Alternative routes to order by disorder},\ }\href
  {https://doi.org/10.1103/PhysRevB.102.220405} {\bibfield  {journal} {\bibinfo
   {journal} {Phys. Rev. B}\ }\textbf {\bibinfo {volume} {102}},\ \bibinfo
  {pages} {220405(R)} (\bibinfo {year} {2020})}\BibitemShut {NoStop}%
\bibitem [{\citenamefont {Reig-i Plessis}\ \emph {et~al.}(2020)\citenamefont
  {Reig-i Plessis}, \citenamefont {Johnson}, \citenamefont {Lu}, \citenamefont
  {Chen}, \citenamefont {Ruff}, \citenamefont {Upton}, \citenamefont
  {Williams}, \citenamefont {Calder}, \citenamefont {Zhou}, \citenamefont
  {Clancy}, \citenamefont {Aczel},\ and\ \citenamefont
  {MacDougall}}]{plessis2020}%
  \BibitemOpen
  \bibfield  {author} {\bibinfo {author} {\bibfnamefont {D.}~\bibnamefont
  {Reig-i Plessis}}, \bibinfo {author} {\bibfnamefont {T.~A.}\ \bibnamefont
  {Johnson}}, \bibinfo {author} {\bibfnamefont {K.}~\bibnamefont {Lu}},
  \bibinfo {author} {\bibfnamefont {Q.}~\bibnamefont {Chen}}, \bibinfo {author}
  {\bibfnamefont {J.~P.~C.}\ \bibnamefont {Ruff}}, \bibinfo {author}
  {\bibfnamefont {M.~H.}\ \bibnamefont {Upton}}, \bibinfo {author}
  {\bibfnamefont {T.~J.}\ \bibnamefont {Williams}}, \bibinfo {author}
  {\bibfnamefont {S.}~\bibnamefont {Calder}}, \bibinfo {author} {\bibfnamefont
  {H.~D.}\ \bibnamefont {Zhou}}, \bibinfo {author} {\bibfnamefont {J.~P.}\
  \bibnamefont {Clancy}}, \bibinfo {author} {\bibfnamefont {A.~A.}\
  \bibnamefont {Aczel}},\ and\ \bibinfo {author} {\bibfnamefont {G.~J.}\
  \bibnamefont {MacDougall}},\ }\bibfield  {title} {\bibinfo {title}
  {Structural, electronic, and magnetic properties of nearly ideal {$J_{\rm
  eff}=\frac12$} iridium halides},\ }\href
  {https://doi.org/10.1103/PhysRevMaterials.4.124407} {\bibfield  {journal}
  {\bibinfo  {journal} {Phys. Rev. Materials}\ }\textbf {\bibinfo {volume}
  {4}},\ \bibinfo {pages} {124407} (\bibinfo {year} {2020})}\BibitemShut
  {NoStop}%
\bibitem [{\citenamefont {Pet{\u r}{\'\i}{\u c}ek}\ \emph
  {et~al.}(2014)\citenamefont {Pet{\u r}{\'\i}{\u c}ek}, \citenamefont {Du{\u
  s}ek},\ and\ \citenamefont {Palatinus}}]{jana2006}%
  \BibitemOpen
  \bibfield  {author} {\bibinfo {author} {\bibfnamefont {V.}~\bibnamefont
  {Pet{\u r}{\'\i}{\u c}ek}}, \bibinfo {author} {\bibfnamefont
  {M.}~\bibnamefont {Du{\u s}ek}},\ and\ \bibinfo {author} {\bibfnamefont
  {L.}~\bibnamefont {Palatinus}},\ }\bibfield  {title} {\bibinfo {title}
  {Crystallographic computing system {JANA2006}: General features},\ }\href
  {https://doi.org/10.1515/zkri-2014-1737} {\bibfield  {journal} {\bibinfo
  {journal} {Z. Krist.}\ }\textbf {\bibinfo {volume} {229}},\ \bibinfo {pages}
  {345} (\bibinfo {year} {2014})}\BibitemShut {NoStop}%
\bibitem [{\citenamefont {K\"uchler}\ \emph {et~al.}(2017)\citenamefont
  {K\"uchler}, \citenamefont {W\"orl}, \citenamefont {Gegenwart}, \citenamefont
  {Berben}, \citenamefont {Bryant},\ and\ \citenamefont
  {Wiedmann}}]{kuechler2017}%
  \BibitemOpen
  \bibfield  {author} {\bibinfo {author} {\bibfnamefont {R.}~\bibnamefont
  {K\"uchler}}, \bibinfo {author} {\bibfnamefont {A.}~\bibnamefont {W\"orl}},
  \bibinfo {author} {\bibfnamefont {P.}~\bibnamefont {Gegenwart}}, \bibinfo
  {author} {\bibfnamefont {M.}~\bibnamefont {Berben}}, \bibinfo {author}
  {\bibfnamefont {B.}~\bibnamefont {Bryant}},\ and\ \bibinfo {author}
  {\bibfnamefont {S.}~\bibnamefont {Wiedmann}},\ }\bibfield  {title} {\bibinfo
  {title} {The world's smallest capacitive dilatometer, for high-resolution
  thermal expansion and magnetostriction in high magnetic fields},\ }\href
  {https://doi.org/10.1063/1.4997073} {\bibfield  {journal} {\bibinfo
  {journal} {Rev. Sci. Instr.}\ }\textbf {\bibinfo {volume} {88}},\ \bibinfo
  {pages} {083903} (\bibinfo {year} {2017})}\BibitemShut {NoStop}%
\bibitem [{\citenamefont {Tsirlin}\ \emph {et~al.}(2009)\citenamefont
  {Tsirlin}, \citenamefont {Schmidt}, \citenamefont {Skourski}, \citenamefont
  {Nath}, \citenamefont {Geibel},\ and\ \citenamefont {Rosner}}]{tsirlin2009}%
  \BibitemOpen
  \bibfield  {author} {\bibinfo {author} {\bibfnamefont {A.~A.}\ \bibnamefont
  {Tsirlin}}, \bibinfo {author} {\bibfnamefont {B.}~\bibnamefont {Schmidt}},
  \bibinfo {author} {\bibfnamefont {Y.}~\bibnamefont {Skourski}}, \bibinfo
  {author} {\bibfnamefont {R.}~\bibnamefont {Nath}}, \bibinfo {author}
  {\bibfnamefont {C.}~\bibnamefont {Geibel}},\ and\ \bibinfo {author}
  {\bibfnamefont {H.}~\bibnamefont {Rosner}},\ }\bibfield  {title} {\bibinfo
  {title} {Exploring the spin-$\frac 12$ frustrated square lattice model with
  high-field magnetization studies},\ }\href
  {https://doi.org/10.1103/PhysRevB.80.132407} {\bibfield  {journal} {\bibinfo
  {journal} {Phys. Rev. B}\ }\textbf {\bibinfo {volume} {80}},\ \bibinfo
  {pages} {132407} (\bibinfo {year} {2009})}\BibitemShut {NoStop}%
\bibitem [{\citenamefont {Koepernik}\ and\ \citenamefont
  {Eschrig}(1999)}]{fplo}%
  \BibitemOpen
  \bibfield  {author} {\bibinfo {author} {\bibfnamefont {K.}~\bibnamefont
  {Koepernik}}\ and\ \bibinfo {author} {\bibfnamefont {H.}~\bibnamefont
  {Eschrig}},\ }\bibfield  {title} {\bibinfo {title} {Full-potential
  nonorthogonal local-orbital minimum-basis band-structure scheme},\ }\href
  {https://doi.org/10.1103/PhysRevB.59.1743} {\bibfield  {journal} {\bibinfo
  {journal} {Phys. Rev. B}\ }\textbf {\bibinfo {volume} {59}},\ \bibinfo
  {pages} {1743} (\bibinfo {year} {1999})}\BibitemShut {NoStop}%
\bibitem [{\citenamefont {Togo}\ and\ \citenamefont {Tanaka}(2015)}]{phonopy}%
  \BibitemOpen
  \bibfield  {author} {\bibinfo {author} {\bibfnamefont {A.}~\bibnamefont
  {Togo}}\ and\ \bibinfo {author} {\bibfnamefont {I.}~\bibnamefont {Tanaka}},\
  }\bibfield  {title} {\bibinfo {title} {First principles phonon calculations
  in materials science},\ }\href
  {https://doi.org/10.1016/j.scriptamat.2015.07.021} {\bibfield  {journal}
  {\bibinfo  {journal} {Scr. Mater.}\ }\textbf {\bibinfo {volume} {108}},\
  \bibinfo {pages} {1} (\bibinfo {year} {2015})}\BibitemShut {NoStop}%
\bibitem [{\citenamefont {Kresse}\ and\ \citenamefont
  {Furthm\"uller}(1996{\natexlab{a}})}]{vasp1}%
  \BibitemOpen
  \bibfield  {author} {\bibinfo {author} {\bibfnamefont {G.}~\bibnamefont
  {Kresse}}\ and\ \bibinfo {author} {\bibfnamefont {J.}~\bibnamefont
  {Furthm\"uller}},\ }\bibfield  {title} {\bibinfo {title} {Efficiency of
  \textit{ab-initio} total energy calculations for metals and semiconductors
  using a plane-wave basis set},\ }\href
  {https://doi.org/http://dx.doi.org/10.1016/0927-0256(96)00008-0} {\bibfield
  {journal} {\bibinfo  {journal} {Computational Materials Science}\ }\textbf
  {\bibinfo {volume} {6}},\ \bibinfo {pages} {15} (\bibinfo {year}
  {1996}{\natexlab{a}})}\BibitemShut {NoStop}%
\bibitem [{\citenamefont {Kresse}\ and\ \citenamefont
  {Furthm\"uller}(1996{\natexlab{b}})}]{vasp2}%
  \BibitemOpen
  \bibfield  {author} {\bibinfo {author} {\bibfnamefont {G.}~\bibnamefont
  {Kresse}}\ and\ \bibinfo {author} {\bibfnamefont {J.}~\bibnamefont
  {Furthm\"uller}},\ }\bibfield  {title} {\bibinfo {title} {Efficient iterative
  schemes for \textit{ab initio} total-energy calculations using a plane-wave
  basis set},\ }\href {https://doi.org/10.1103/PhysRevB.54.11169} {\bibfield
  {journal} {\bibinfo  {journal} {Phys. Rev. B}\ }\textbf {\bibinfo {volume}
  {54}},\ \bibinfo {pages} {11169} (\bibinfo {year}
  {1996}{\natexlab{b}})}\BibitemShut {NoStop}%
\bibitem [{\citenamefont {Perdew}\ \emph {et~al.}(1996)\citenamefont {Perdew},
  \citenamefont {Burke},\ and\ \citenamefont {Ernzerhof}}]{pbe96}%
  \BibitemOpen
  \bibfield  {author} {\bibinfo {author} {\bibfnamefont {J.~P.}\ \bibnamefont
  {Perdew}}, \bibinfo {author} {\bibfnamefont {K.}~\bibnamefont {Burke}},\ and\
  \bibinfo {author} {\bibfnamefont {M.}~\bibnamefont {Ernzerhof}},\ }\bibfield
  {title} {\bibinfo {title} {Generalized gradient approximation made simple},\
  }\href {https://doi.org/10.1103/PhysRevLett.77.3865} {\bibfield  {journal}
  {\bibinfo  {journal} {Phys. Rev. Lett.}\ }\textbf {\bibinfo {volume} {77}},\
  \bibinfo {pages} {3865} (\bibinfo {year} {1996})}\BibitemShut {NoStop}%
\bibitem [{\citenamefont {R\"ossler}\ and\ \citenamefont
  {Winter}(1977)}]{rossler1977}%
  \BibitemOpen
  \bibfield  {author} {\bibinfo {author} {\bibfnamefont {K.}~\bibnamefont
  {R\"ossler}}\ and\ \bibinfo {author} {\bibfnamefont {J.}~\bibnamefont
  {Winter}},\ }\bibfield  {title} {\bibinfo {title} {Influence of $d$-electron
  configuration on phase transitions in {A$_2$MX$_6$} (Hexahalometallates
  IV)},\ }\href {https://doi.org/10.1016/0009-2614(77)80653-2} {\bibfield
  {journal} {\bibinfo  {journal} {Chem. Phys. Lett.}\ }\textbf {\bibinfo
  {volume} {46}},\ \bibinfo {pages} {566} (\bibinfo {year} {1977})}\BibitemShut
  {NoStop}%
\bibitem [{\citenamefont {Gubanov}\ \emph {et~al.}(2002)\citenamefont
  {Gubanov}, \citenamefont {Gromilov}, \citenamefont {Korenev}, \citenamefont
  {Venediktov},\ and\ \citenamefont {Asanov}}]{gubanov2002}%
  \BibitemOpen
  \bibfield  {author} {\bibinfo {author} {\bibfnamefont {A.~I.}\ \bibnamefont
  {Gubanov}}, \bibinfo {author} {\bibfnamefont {S.~A.}\ \bibnamefont
  {Gromilov}}, \bibinfo {author} {\bibfnamefont {S.~V.}\ \bibnamefont
  {Korenev}}, \bibinfo {author} {\bibfnamefont {A.~B.}\ \bibnamefont
  {Venediktov}},\ and\ \bibinfo {author} {\bibfnamefont {I.~P.}\ \bibnamefont
  {Asanov}},\ }\bibfield  {title} {\bibinfo {title} {Synthesis and study of
  potassium hexabromoiridate{(IV)}},\ }\href
  {https://doi.org/10.1023/A:1021638429901} {\bibfield  {journal} {\bibinfo
  {journal} {Russ. J. Coord. Chem.}\ }\textbf {\bibinfo {volume} {28}},\
  \bibinfo {pages} {864} (\bibinfo {year} {2002})}\BibitemShut {NoStop}%
\bibitem [{\citenamefont {Abriel}(1982)}]{abriel1982}%
  \BibitemOpen
  \bibfield  {author} {\bibinfo {author} {\bibfnamefont {W.}~\bibnamefont
  {Abriel}},\ }\bibfield  {title} {\bibinfo {title} {Crystal structure and
  phase transition of {Rb$_2$TeI$_6$}},\ }\href
  {https://doi.org/10.1016/0025-5408(82)90171-4} {\bibfield  {journal}
  {\bibinfo  {journal} {Mater. Res. Bull.}\ }\textbf {\bibinfo {volume} {17}},\
  \bibinfo {pages} {1341} (\bibinfo {year} {1982})}\BibitemShut {NoStop}%
\bibitem [{\citenamefont {Abriel}(1984)}]{abriel1984}%
  \BibitemOpen
  \bibfield  {author} {\bibinfo {author} {\bibfnamefont {W.}~\bibnamefont
  {Abriel}},\ }\bibfield  {title} {\bibinfo {title} {Polymorphism and phase
  transitions of {K$_2$TeBr$_6$}: A single crystal investigation of the high
  temperature phases {(383--463 K)}},\ }\href
  {https://doi.org/10.1016/0025-5408(84)90172-7} {\bibfield  {journal}
  {\bibinfo  {journal} {Mater. Res. Bull.}\ }\textbf {\bibinfo {volume} {19}},\
  \bibinfo {pages} {313} (\bibinfo {year} {1984})}\BibitemShut {NoStop}%
\bibitem [{\citenamefont {Abrahams}\ \emph {et~al.}(1984)\citenamefont
  {Abrahams}, \citenamefont {Ihringer}, \citenamefont {Marsh},\ and\
  \citenamefont {Nassau}}]{abrahams1984}%
  \BibitemOpen
  \bibfield  {author} {\bibinfo {author} {\bibfnamefont {S.~C.}\ \bibnamefont
  {Abrahams}}, \bibinfo {author} {\bibfnamefont {J.}~\bibnamefont {Ihringer}},
  \bibinfo {author} {\bibfnamefont {P.}~\bibnamefont {Marsh}},\ and\ \bibinfo
  {author} {\bibfnamefont {K.}~\bibnamefont {Nassau}},\ }\bibfield  {title}
  {\bibinfo {title} {Phase transition at {434 K}, independent strain coupling
  in second transition at {400 K}, and thermal expansivity in ferroelastic
  {K$_2$TeBr$_6$}},\ }\href {https://doi.org/10.1063/1.447832} {\bibfield
  {journal} {\bibinfo  {journal} {J. Chem. Phys.}\ }\textbf {\bibinfo {volume}
  {81}},\ \bibinfo {pages} {2082} (\bibinfo {year} {1984})}\BibitemShut
  {NoStop}%
\bibitem [{sup()}]{suppl}%
  \BibitemOpen
  \href@noop {} {}\bibinfo {note} {See Supplemental Material for additional XRD
  data and structure refinements, as well as additional magnetization data,
  exchange tensors, and details of DFT calculations.}\BibitemShut {Stop}%
\bibitem [{\citenamefont {Lynn}\ \emph {et~al.}(1978)\citenamefont {Lynn},
  \citenamefont {Patterson}, \citenamefont {Shirane},\ and\ \citenamefont
  {Wheeler}}]{lynn1978}%
  \BibitemOpen
  \bibfield  {author} {\bibinfo {author} {\bibfnamefont {J.~W.}\ \bibnamefont
  {Lynn}}, \bibinfo {author} {\bibfnamefont {H.~H.}\ \bibnamefont {Patterson}},
  \bibinfo {author} {\bibfnamefont {G.}~\bibnamefont {Shirane}},\ and\ \bibinfo
  {author} {\bibfnamefont {R.~G.}\ \bibnamefont {Wheeler}},\ }\bibfield
  {title} {\bibinfo {title} {Soft rotary mode and structural phase transitions
  in {K$_2$ReCl$_6$}},\ }\href {https://doi.org/10.1016/0038-1098(78)90192-8}
  {\bibfield  {journal} {\bibinfo  {journal} {Solid State Comm.}\ }\textbf
  {\bibinfo {volume} {27}},\ \bibinfo {pages} {859} (\bibinfo {year}
  {1978})}\BibitemShut {NoStop}%
\bibitem [{\citenamefont {Mintz}\ \emph {et~al.}(1979)\citenamefont {Mintz},
  \citenamefont {Armstrong}, \citenamefont {Powell},\ and\ \citenamefont
  {Buyers}}]{mintz1979}%
  \BibitemOpen
  \bibfield  {author} {\bibinfo {author} {\bibfnamefont {D.}~\bibnamefont
  {Mintz}}, \bibinfo {author} {\bibfnamefont {R.~L.}\ \bibnamefont
  {Armstrong}}, \bibinfo {author} {\bibfnamefont {B.~M.}\ \bibnamefont
  {Powell}},\ and\ \bibinfo {author} {\bibfnamefont {W.~J.~L.}\ \bibnamefont
  {Buyers}},\ }\bibfield  {title} {\bibinfo {title} {Soft rotary mode in the
  antifluorite crystal {K$_2$OsCl$_6$}},\ }\href
  {https://doi.org/10.1103/PhysRevB.19.448} {\bibfield  {journal} {\bibinfo
  {journal} {Phys. Rev. B}\ }\textbf {\bibinfo {volume} {19}},\ \bibinfo
  {pages} {448} (\bibinfo {year} {1979})}\BibitemShut {NoStop}%
\bibitem [{\citenamefont {Sutton}\ \emph {et~al.}(1983)\citenamefont {Sutton},
  \citenamefont {Armstrong}, \citenamefont {Powell},\ and\ \citenamefont
  {Buyers}}]{sutton1983}%
  \BibitemOpen
  \bibfield  {author} {\bibinfo {author} {\bibfnamefont {M.}~\bibnamefont
  {Sutton}}, \bibinfo {author} {\bibfnamefont {R.~L.}\ \bibnamefont
  {Armstrong}}, \bibinfo {author} {\bibfnamefont {B.~M.}\ \bibnamefont
  {Powell}},\ and\ \bibinfo {author} {\bibfnamefont {W.~J.~L.}\ \bibnamefont
  {Buyers}},\ }\bibfield  {title} {\bibinfo {title} {Lattice dynamics and phase
  transitions in antifluorite crystals: {K$_2$OsCl$_6$}},\ }\href
  {https://doi.org/10.1103/PhysRevB.27.380} {\bibfield  {journal} {\bibinfo
  {journal} {Phys. Rev. B}\ }\textbf {\bibinfo {volume} {27}},\ \bibinfo
  {pages} {380} (\bibinfo {year} {1983})}\BibitemShut {NoStop}%
\bibitem [{\citenamefont {Novotny}\ \emph {et~al.}(1977)\citenamefont
  {Novotny}, \citenamefont {Martin}, \citenamefont {Armstrong},\ and\
  \citenamefont {Meincke}}]{novotny1977}%
  \BibitemOpen
  \bibfield  {author} {\bibinfo {author} {\bibfnamefont {V.}~\bibnamefont
  {Novotny}}, \bibinfo {author} {\bibfnamefont {C.~A.}\ \bibnamefont {Martin}},
  \bibinfo {author} {\bibfnamefont {R.~L.}\ \bibnamefont {Armstrong}},\ and\
  \bibinfo {author} {\bibfnamefont {P.~P.~M.}\ \bibnamefont {Meincke}},\
  }\bibfield  {title} {\bibinfo {title} {Thermodynamical properties of
  {K$_2$OsCl$_6$} and {K$_2$ReCl$_6$} at low temperatures and near their
  structural phase transitions},\ }\href
  {https://doi.org/10.1103/PhysRevB.15.382} {\bibfield  {journal} {\bibinfo
  {journal} {Phys. Rev. B}\ }\textbf {\bibinfo {volume} {15}},\ \bibinfo
  {pages} {382} (\bibinfo {year} {1977})}\BibitemShut {NoStop}%
\bibitem [{\citenamefont {Armstrong}\ \emph {et~al.}(1978)\citenamefont
  {Armstrong}, \citenamefont {Mintz}, \citenamefont {Powell},\ and\
  \citenamefont {Buyers}}]{armstrong1978}%
  \BibitemOpen
  \bibfield  {author} {\bibinfo {author} {\bibfnamefont {R.~L.}\ \bibnamefont
  {Armstrong}}, \bibinfo {author} {\bibfnamefont {D.}~\bibnamefont {Mintz}},
  \bibinfo {author} {\bibfnamefont {B.~M.}\ \bibnamefont {Powell}},\ and\
  \bibinfo {author} {\bibfnamefont {W.~J.~L.}\ \bibnamefont {Buyers}},\
  }\bibfield  {title} {\bibinfo {title} {Ferrorotative transition in the
  antifluorite crystal {K$_2$OsCl$_6$}},\ }\href
  {https://doi.org/10.1103/PhysRevB.17.1260} {\bibfield  {journal} {\bibinfo
  {journal} {Phys. Rev. B}\ }\textbf {\bibinfo {volume} {17}},\ \bibinfo
  {pages} {1260} (\bibinfo {year} {1978})}\BibitemShut {NoStop}%
\bibitem [{\citenamefont {Willemsen}\ \emph {et~al.}(1977)\citenamefont
  {Willemsen}, \citenamefont {Martin}, \citenamefont {Meincke},\ and\
  \citenamefont {Armstrong}}]{willemsen1977}%
  \BibitemOpen
  \bibfield  {author} {\bibinfo {author} {\bibfnamefont {H.~W.}\ \bibnamefont
  {Willemsen}}, \bibinfo {author} {\bibfnamefont {C.~A.}\ \bibnamefont
  {Martin}}, \bibinfo {author} {\bibfnamefont {P.~P.~M.}\ \bibnamefont
  {Meincke}},\ and\ \bibinfo {author} {\bibfnamefont {R.~L.}\ \bibnamefont
  {Armstrong}},\ }\bibfield  {title} {\bibinfo {title} {Thermal-expansion study
  of the displacive phase transitions in {K$_2$ReCl$_6$} and {K$_2$OsCl$_6$}},\
  }\href {https://doi.org/10.1103/PhysRevB.16.2283} {\bibfield  {journal}
  {\bibinfo  {journal} {Phys. Rev. B}\ }\textbf {\bibinfo {volume} {16}},\
  \bibinfo {pages} {2283} (\bibinfo {year} {1977})}\BibitemShut {NoStop}%
\bibitem [{\citenamefont {Cowley}(1980)}]{cowley1980}%
  \BibitemOpen
  \bibfield  {author} {\bibinfo {author} {\bibfnamefont {R.~A.}\ \bibnamefont
  {Cowley}},\ }\bibfield  {title} {\bibinfo {title} {Structural phase
  transitions I. Landau theory},\ }\href
  {https://doi.org/10.1080/00018738000101346} {\bibfield  {journal} {\bibinfo
  {journal} {Adv. Phys.}\ }\textbf {\bibinfo {volume} {29}},\ \bibinfo {pages}
  {1} (\bibinfo {year} {1980})}\BibitemShut {NoStop}%
\bibitem [{\citenamefont {{Van Driel}}\ \emph {et~al.}(1972)\citenamefont {{Van
  Driel}}, \citenamefont {Wiszniewska}, \citenamefont {Moores},\ and\
  \citenamefont {Armstrong}}]{vandriel1972}%
  \BibitemOpen
  \bibfield  {author} {\bibinfo {author} {\bibfnamefont {H.~M.}\ \bibnamefont
  {{Van Driel}}}, \bibinfo {author} {\bibfnamefont {M.}~\bibnamefont
  {Wiszniewska}}, \bibinfo {author} {\bibfnamefont {B.~M.}\ \bibnamefont
  {Moores}},\ and\ \bibinfo {author} {\bibfnamefont {R.~L.}\ \bibnamefont
  {Armstrong}},\ }\bibfield  {title} {\bibinfo {title} {Softening of the rotary
  lattice mode in {K$_2$PtBr$_6$} as detected by nuclear quadrupole
  resonance},\ }\href {https://doi.org/10.1103/PhysRevB.6.1596} {\bibfield
  {journal} {\bibinfo  {journal} {Phys. Rev. B}\ }\textbf {\bibinfo {volume}
  {6}},\ \bibinfo {pages} {1596} (\bibinfo {year} {1972})}\BibitemShut
  {NoStop}%
\bibitem [{\citenamefont {Widmann}\ \emph {et~al.}(2019)\citenamefont
  {Widmann}, \citenamefont {Tsurkan}, \citenamefont {Prishchenko},
  \citenamefont {Mazurenko}, \citenamefont {Tsirlin},\ and\ \citenamefont
  {Loidl}}]{widmann2019}%
  \BibitemOpen
  \bibfield  {author} {\bibinfo {author} {\bibfnamefont {S.}~\bibnamefont
  {Widmann}}, \bibinfo {author} {\bibfnamefont {V.}~\bibnamefont {Tsurkan}},
  \bibinfo {author} {\bibfnamefont {D.~A.}\ \bibnamefont {Prishchenko}},
  \bibinfo {author} {\bibfnamefont {V.~G.}\ \bibnamefont {Mazurenko}}, \bibinfo
  {author} {\bibfnamefont {A.~A.}\ \bibnamefont {Tsirlin}},\ and\ \bibinfo
  {author} {\bibfnamefont {A.}~\bibnamefont {Loidl}},\ }\bibfield  {title}
  {\bibinfo {title} {Thermodynamic evidence of fractionalized excitations in
  {$\alpha$-RuCl$_3$}},\ }\href {https://doi.org/10.1103/PhysRevB.99.094415}
  {\bibfield  {journal} {\bibinfo  {journal} {Phys. Rev. B}\ }\textbf {\bibinfo
  {volume} {99}},\ \bibinfo {pages} {094415} (\bibinfo {year}
  {2019})}\BibitemShut {NoStop}%
\bibitem [{\citenamefont {Huang}\ \emph {et~al.}(1994)\citenamefont {Huang},
  \citenamefont {Soubeyroux}, \citenamefont {Chmaissem}, \citenamefont {{Natali
  Sora}}, \citenamefont {Santoro}, \citenamefont {Cava}, \citenamefont
  {Krajewski},\ and\ \citenamefont {{Peck Jr.}}}]{huang1994}%
  \BibitemOpen
  \bibfield  {author} {\bibinfo {author} {\bibfnamefont {Q.}~\bibnamefont
  {Huang}}, \bibinfo {author} {\bibfnamefont {J.~L.}\ \bibnamefont
  {Soubeyroux}}, \bibinfo {author} {\bibfnamefont {O.}~\bibnamefont
  {Chmaissem}}, \bibinfo {author} {\bibfnamefont {I.}~\bibnamefont {{Natali
  Sora}}}, \bibinfo {author} {\bibfnamefont {A.}~\bibnamefont {Santoro}},
  \bibinfo {author} {\bibfnamefont {R.~J.}\ \bibnamefont {Cava}}, \bibinfo
  {author} {\bibfnamefont {J.~J.}\ \bibnamefont {Krajewski}},\ and\ \bibinfo
  {author} {\bibfnamefont {W.~F.}\ \bibnamefont {{Peck Jr.}}},\ }\bibfield
  {title} {\bibinfo {title} {Neutron powder diffraction study of the crystal
  structures of {Sr$_2$RuO$_4$} and {Sr$_2$IrO$_4$} at room temperature and at
  {10 K}},\ }\href {https://doi.org/10.1006/jssc.1994.1316} {\bibfield
  {journal} {\bibinfo  {journal} {J. Solid State Chem.}\ }\textbf {\bibinfo
  {volume} {112}},\ \bibinfo {pages} {355} (\bibinfo {year}
  {1994})}\BibitemShut {NoStop}%
\bibitem [{\citenamefont {Nakatsuka}\ \emph {et~al.}(2015)\citenamefont
  {Nakatsuka}, \citenamefont {Sugiyama}, \citenamefont {Yoneda}, \citenamefont
  {Fujiwara},\ and\ \citenamefont {Yoshiasa}}]{nakatsuka2015}%
  \BibitemOpen
  \bibfield  {author} {\bibinfo {author} {\bibfnamefont {A.}~\bibnamefont
  {Nakatsuka}}, \bibinfo {author} {\bibfnamefont {K.}~\bibnamefont {Sugiyama}},
  \bibinfo {author} {\bibfnamefont {A.}~\bibnamefont {Yoneda}}, \bibinfo
  {author} {\bibfnamefont {K.}~\bibnamefont {Fujiwara}},\ and\ \bibinfo
  {author} {\bibfnamefont {A.}~\bibnamefont {Yoshiasa}},\ }\bibfield  {title}
  {\bibinfo {title} {Crystal structure of post-perovskite-type {CaIrO$_3$}
  reinvestigated: new insights into atomic thermal vibration behaviors},\
  }\href {https://doi.org/10.1107/S2056989015015649} {\bibfield  {journal}
  {\bibinfo  {journal} {Acta Cryst.}\ }\textbf {\bibinfo {volume} {E71}},\
  \bibinfo {pages} {1109} (\bibinfo {year} {2015})}\BibitemShut {NoStop}%
\bibitem [{\citenamefont {{Moretti Sala}}\ \emph {et~al.}(2014)\citenamefont
  {{Moretti Sala}}, \citenamefont {Ohgushi}, \citenamefont {Al-Zein},
  \citenamefont {Hirata}, \citenamefont {Monaco},\ and\ \citenamefont
  {Krisch}}]{moretti2014}%
  \BibitemOpen
  \bibfield  {author} {\bibinfo {author} {\bibfnamefont {M.}~\bibnamefont
  {{Moretti Sala}}}, \bibinfo {author} {\bibfnamefont {K.}~\bibnamefont
  {Ohgushi}}, \bibinfo {author} {\bibfnamefont {A.}~\bibnamefont {Al-Zein}},
  \bibinfo {author} {\bibfnamefont {Y.}~\bibnamefont {Hirata}}, \bibinfo
  {author} {\bibfnamefont {G.}~\bibnamefont {Monaco}},\ and\ \bibinfo {author}
  {\bibfnamefont {M.}~\bibnamefont {Krisch}},\ }\bibfield  {title} {\bibinfo
  {title} {{CaIrO$_3$}: A spin-orbit {Mott} insulator beyond the $j_{\rm
  eff}=\frac12$ ground state},\ }\href
  {https://doi.org/10.1103/PhysRevLett.112.176402} {\bibfield  {journal}
  {\bibinfo  {journal} {Phys. Rev. Lett.}\ }\textbf {\bibinfo {volume} {112}},\
  \bibinfo {pages} {176402} (\bibinfo {year} {2014})}\BibitemShut {NoStop}%
\bibitem [{Note2()}]{Note2}%
  \BibitemOpen
  \bibinfo {note} {Here, we neglect the weak splitting of $3-4$\protect
  \tmspace +\thinmuskip {.1667em}meV that appears between the doubly-degenerate
  levels upon orthorhombic deformation.}\BibitemShut {Stop}%
\bibitem [{\citenamefont {Bogdanov}\ \emph {et~al.}(2015)\citenamefont
  {Bogdanov}, \citenamefont {Katukuri}, \citenamefont {Romh\'anyi},
  \citenamefont {Yushankhai}, \citenamefont {Kataev}, \citenamefont
  {B\"uchner}, \citenamefont {{van den Brink}},\ and\ \citenamefont
  {Hozoi}}]{bogdanov2015}%
  \BibitemOpen
  \bibfield  {author} {\bibinfo {author} {\bibfnamefont {N.~A.}\ \bibnamefont
  {Bogdanov}}, \bibinfo {author} {\bibfnamefont {V.~M.}\ \bibnamefont
  {Katukuri}}, \bibinfo {author} {\bibfnamefont {J.}~\bibnamefont
  {Romh\'anyi}}, \bibinfo {author} {\bibfnamefont {V.}~\bibnamefont
  {Yushankhai}}, \bibinfo {author} {\bibfnamefont {V.}~\bibnamefont {Kataev}},
  \bibinfo {author} {\bibfnamefont {B.}~\bibnamefont {B\"uchner}}, \bibinfo
  {author} {\bibfnamefont {J.}~\bibnamefont {{van den Brink}}},\ and\ \bibinfo
  {author} {\bibfnamefont {L.}~\bibnamefont {Hozoi}},\ }\bibfield  {title}
  {\bibinfo {title} {Orbital reconstruction in nonpolar tetravalent
  transition-metal oxide layers},\ }\href {https://doi.org/10.1038/ncomms8306}
  {\bibfield  {journal} {\bibinfo  {journal} {Nature Comm.}\ }\textbf {\bibinfo
  {volume} {6}},\ \bibinfo {pages} {7306} (\bibinfo {year} {2015})}\BibitemShut
  {NoStop}%
\bibitem [{\citenamefont {Lynn}\ \emph {et~al.}(1976)\citenamefont {Lynn},
  \citenamefont {Shirane},\ and\ \citenamefont {Blume}}]{lynn1976}%
  \BibitemOpen
  \bibfield  {author} {\bibinfo {author} {\bibfnamefont {J.~W.}\ \bibnamefont
  {Lynn}}, \bibinfo {author} {\bibfnamefont {G.}~\bibnamefont {Shirane}},\ and\
  \bibinfo {author} {\bibfnamefont {M.}~\bibnamefont {Blume}},\ }\bibfield
  {title} {\bibinfo {title} {Covalency effects in the magnetic form factor of
  {Ir} in {K$_2$IrCl$_6$}},\ }\href
  {https://doi.org/10.1103/PhysRevLett.37.154} {\bibfield  {journal} {\bibinfo
  {journal} {Phys. Rev. Lett.}\ }\textbf {\bibinfo {volume} {37}},\ \bibinfo
  {pages} {154} (\bibinfo {year} {1976})}\BibitemShut {NoStop}%
\bibitem [{\citenamefont {Bain}\ and\ \citenamefont {Berry}(2008)}]{bain2008}%
  \BibitemOpen
  \bibfield  {author} {\bibinfo {author} {\bibfnamefont {G.~A.}\ \bibnamefont
  {Bain}}\ and\ \bibinfo {author} {\bibfnamefont {J.~F.}\ \bibnamefont
  {Berry}},\ }\bibfield  {title} {\bibinfo {title} {Diamagnetic corrections and
  {Pascal's} constants},\ }\href {https://doi.org/10.1021/ed085p532} {\bibfield
   {journal} {\bibinfo  {journal} {J. Chem. Education}\ }\textbf {\bibinfo
  {volume} {85}},\ \bibinfo {pages} {532} (\bibinfo {year} {2008})}\BibitemShut
  {NoStop}%
\bibitem [{\citenamefont {Majumder}\ \emph {et~al.}(2019)\citenamefont
  {Majumder}, \citenamefont {Freund}, \citenamefont {Dey}, \citenamefont
  {Prinz-Zwick}, \citenamefont {B\"uttgen}, \citenamefont {Skourski},
  \citenamefont {Jesche}, \citenamefont {Tsirlin},\ and\ \citenamefont
  {Gegenwart}}]{majumder2019}%
  \BibitemOpen
  \bibfield  {author} {\bibinfo {author} {\bibfnamefont {M.}~\bibnamefont
  {Majumder}}, \bibinfo {author} {\bibfnamefont {F.}~\bibnamefont {Freund}},
  \bibinfo {author} {\bibfnamefont {T.}~\bibnamefont {Dey}}, \bibinfo {author}
  {\bibfnamefont {M.}~\bibnamefont {Prinz-Zwick}}, \bibinfo {author}
  {\bibfnamefont {N.}~\bibnamefont {B\"uttgen}}, \bibinfo {author}
  {\bibfnamefont {Y.}~\bibnamefont {Skourski}}, \bibinfo {author}
  {\bibfnamefont {A.}~\bibnamefont {Jesche}}, \bibinfo {author} {\bibfnamefont
  {A.~A.}\ \bibnamefont {Tsirlin}},\ and\ \bibinfo {author} {\bibfnamefont
  {P.}~\bibnamefont {Gegenwart}},\ }\bibfield  {title} {\bibinfo {title}
  {Anisotropic temperature-field phase diagram of single crystalline
  {$\beta$-Li$_2$IrO$_3$}: Magnetization, specific heat, and {$^7$Li NMR}
  study},\ }\href {https://doi.org/10.1103/PhysRevMaterials.3.074408}
  {\bibfield  {journal} {\bibinfo  {journal} {Phys. Rev. Materials}\ }\textbf
  {\bibinfo {volume} {3}},\ \bibinfo {pages} {074408} (\bibinfo {year}
  {2019})}\BibitemShut {NoStop}%
\bibitem [{Note3()}]{Note3}%
  \BibitemOpen
  \bibinfo {note} {Specific heat does not show thermal hysteresis at any of the
  transitions because of long heat pulses used in the relaxation
  method.}\BibitemShut {Stop}%
\bibitem [{\citenamefont {Cao}\ \emph {et~al.}(2013)\citenamefont {Cao},
  \citenamefont {Subedi}, \citenamefont {Calder}, \citenamefont {Yan},
  \citenamefont {Yi}, \citenamefont {Gai}, \citenamefont {Poudel},
  \citenamefont {Singh}, \citenamefont {Lumsden}, \citenamefont {Christianson},
  \citenamefont {Sales},\ and\ \citenamefont {Mandrus}}]{cao2013}%
  \BibitemOpen
  \bibfield  {author} {\bibinfo {author} {\bibfnamefont {G.}~\bibnamefont
  {Cao}}, \bibinfo {author} {\bibfnamefont {A.}~\bibnamefont {Subedi}},
  \bibinfo {author} {\bibfnamefont {S.}~\bibnamefont {Calder}}, \bibinfo
  {author} {\bibfnamefont {J.-Q.}\ \bibnamefont {Yan}}, \bibinfo {author}
  {\bibfnamefont {J.}~\bibnamefont {Yi}}, \bibinfo {author} {\bibfnamefont
  {Z.}~\bibnamefont {Gai}}, \bibinfo {author} {\bibfnamefont {L.}~\bibnamefont
  {Poudel}}, \bibinfo {author} {\bibfnamefont {D.~J.}\ \bibnamefont {Singh}},
  \bibinfo {author} {\bibfnamefont {M.~D.}\ \bibnamefont {Lumsden}}, \bibinfo
  {author} {\bibfnamefont {A.~D.}\ \bibnamefont {Christianson}}, \bibinfo
  {author} {\bibfnamefont {B.~C.}\ \bibnamefont {Sales}},\ and\ \bibinfo
  {author} {\bibfnamefont {D.}~\bibnamefont {Mandrus}},\ }\bibfield  {title}
  {\bibinfo {title} {Magnetism and electronic structure of {La$_2$ZnIrO$_6$}
  and {La$_2$MgIrO$_6$}: Candidate {$J_{\rm eff}=\frac12$} {Mott} insulators},\
  }\href {https://doi.org/10.1103/PhysRevB.87.155136} {\bibfield  {journal}
  {\bibinfo  {journal} {Phys. Rev. B}\ }\textbf {\bibinfo {volume} {87}},\
  \bibinfo {pages} {155136} (\bibinfo {year} {2013})}\BibitemShut {NoStop}%
\bibitem [{\citenamefont {Rau}\ \emph {et~al.}(2014)\citenamefont {Rau},
  \citenamefont {Lee},\ and\ \citenamefont {Kee}}]{rau2014}%
  \BibitemOpen
  \bibfield  {author} {\bibinfo {author} {\bibfnamefont {J.~G.}\ \bibnamefont
  {Rau}}, \bibinfo {author} {\bibfnamefont {E.~K.-H.}\ \bibnamefont {Lee}},\
  and\ \bibinfo {author} {\bibfnamefont {H.-Y.}\ \bibnamefont {Kee}},\
  }\bibfield  {title} {\bibinfo {title} {Generic spin model for the honeycomb
  iridates beyond the {Kitaev} limit},\ }\href
  {https://doi.org/10.1103/PhysRevLett.112.077204} {\bibfield  {journal}
  {\bibinfo  {journal} {Phys. Rev. Lett.}\ }\textbf {\bibinfo {volume} {112}},\
  \bibinfo {pages} {077204} (\bibinfo {year} {2014})}\BibitemShut {NoStop}%
\bibitem [{\citenamefont {Winter}\ \emph {et~al.}(2016)\citenamefont {Winter},
  \citenamefont {Li}, \citenamefont {Jeschke},\ and\ \citenamefont
  {Valent{\'\i}}}]{winter2016}%
  \BibitemOpen
  \bibfield  {author} {\bibinfo {author} {\bibfnamefont {S.~M.}\ \bibnamefont
  {Winter}}, \bibinfo {author} {\bibfnamefont {Y.}~\bibnamefont {Li}}, \bibinfo
  {author} {\bibfnamefont {H.~O.}\ \bibnamefont {Jeschke}},\ and\ \bibinfo
  {author} {\bibfnamefont {R.}~\bibnamefont {Valent{\'\i}}},\ }\bibfield
  {title} {\bibinfo {title} {Challenges in design of {Kitaev} materials:
  Magnetic interactions from competing energy scales},\ }\href
  {https://doi.org/10.1103/PhysRevB.93.214431} {\bibfield  {journal} {\bibinfo
  {journal} {Phys. Rev. B}\ }\textbf {\bibinfo {volume} {93}},\ \bibinfo
  {pages} {214431} (\bibinfo {year} {2016})}\BibitemShut {NoStop}%
\bibitem [{\citenamefont {{Ter Haar}}\ and\ \citenamefont
  {Lines}(1962)}]{haar1962}%
  \BibitemOpen
  \bibfield  {author} {\bibinfo {author} {\bibfnamefont {D.}~\bibnamefont {{Ter
  Haar}}}\ and\ \bibinfo {author} {\bibfnamefont {M.}~\bibnamefont {Lines}},\
  }\bibfield  {title} {\bibinfo {title} {A spin-wave theory of anisotropic
  antiferromagnetica},\ }\href {https://doi.org/10.1098/rsta.1962.0008}
  {\bibfield  {journal} {\bibinfo  {journal} {Phil. Trans. Royal Soc. A}\
  }\textbf {\bibinfo {volume} {255}},\ \bibinfo {pages} {1} (\bibinfo {year}
  {1962})}\BibitemShut {NoStop}%
\end{thebibliography}
%

\clearpage

\begin{widetext}
\begin{center}
\large{\textbf{\textit{Supplemental Material}\smallskip\\
Toward cubic symmetry for Ir$^{4+}$: structure and magnetism of the antifluorite K$_2$IrBr$_6$}}
\end{center}
\end{widetext}

\renewcommand{\thefigure}{S\arabic{figure}}
\renewcommand{\thetable}{S\arabic{table}}
\renewcommand{\theequation}{S\arabic{equation}}
\setcounter{figure}{0}
\setcounter{equation}{0}

\subsection{Crystal structure}

\begin{figure}[htb]
	\centering
	\includegraphics[scale=0.6]{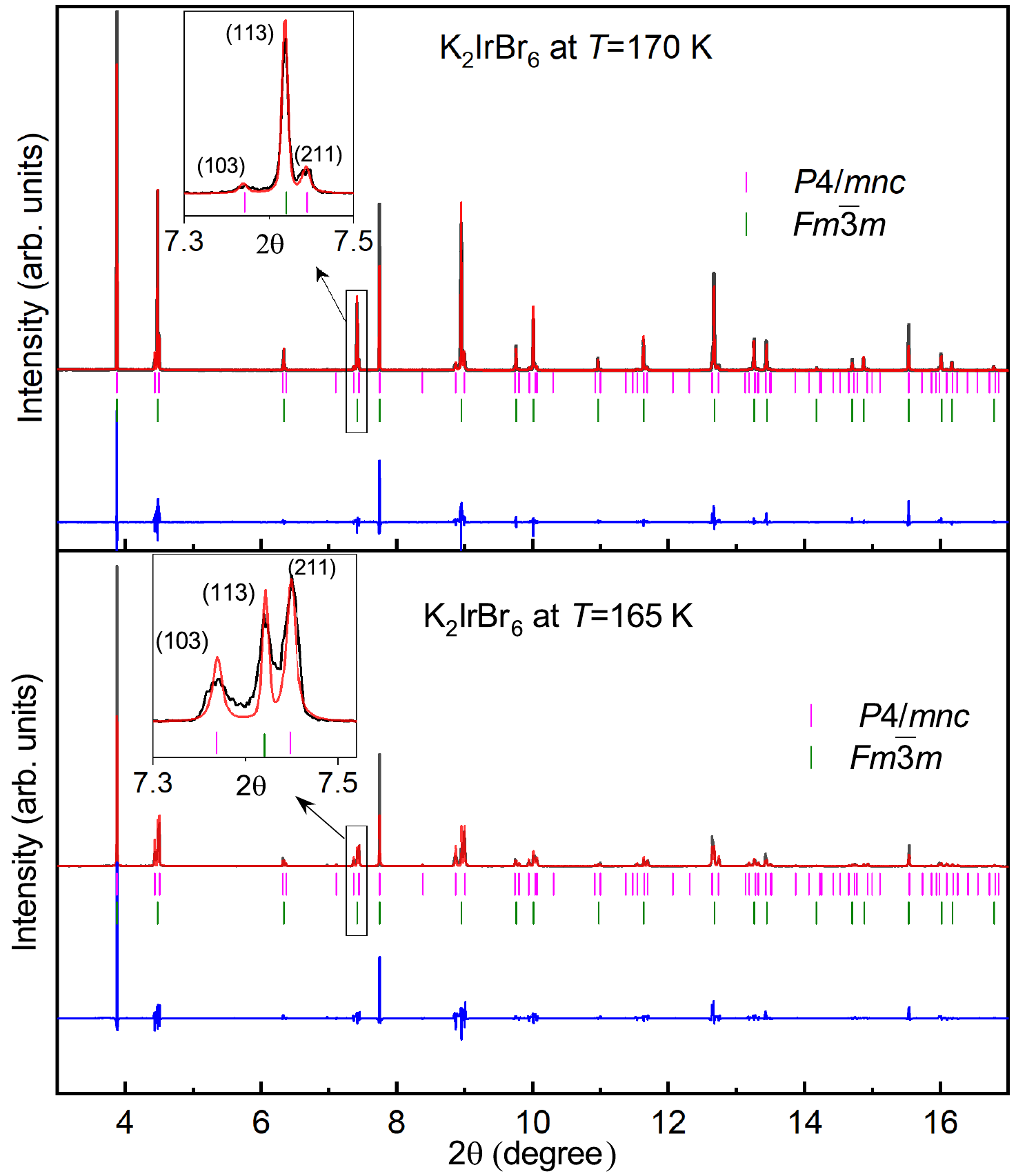}
	\caption{Refined synchrotron powder X-ray diffraction patterns of K$_2$IrBr$_6$ in the temperature range of phase coexistence upon the $\alpha-\beta$ transition.}\label{Fig.S2}
\end{figure}

In Fig.~\ref{Fig.S2}, we show structure refinements in the vicinity of the $\alpha-\beta$ transition to highlight phase coexistence at 165 and 170\,K.

\begin{figure}[t]
	\centering
	\includegraphics[scale=0.5]{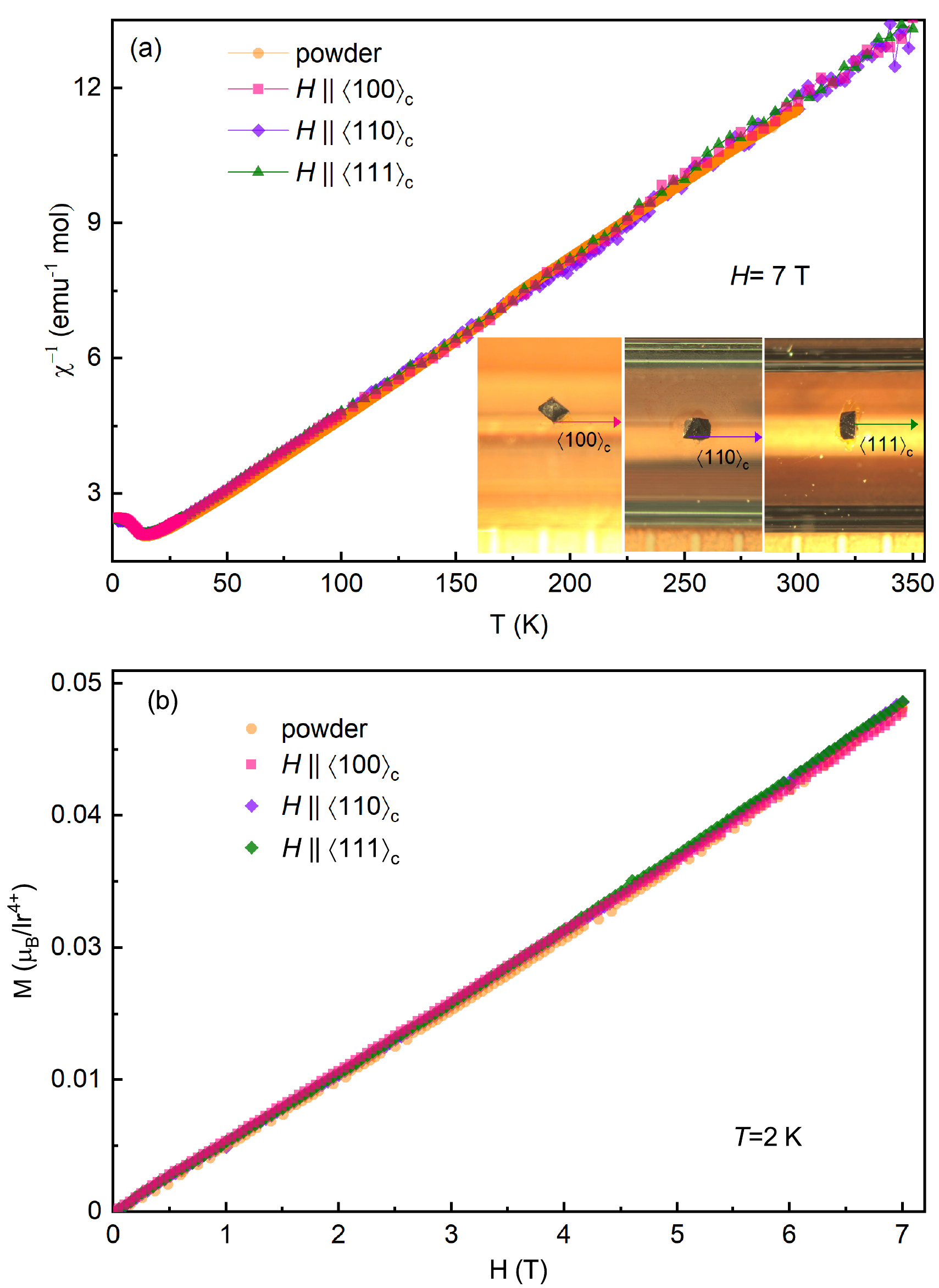}
	\caption{Temperature dependence of the inverse susceptibility (a), and field dependence of the magnetization at 2\,K (b) measured on a 0.29\,mg K$_2$IrBr$_6$ single crystal in the field $\mu_0H=7$\,T applied along $\langle 100\rangle_c$, $\langle 110\rangle_c$, and $\langle 111\rangle_c$ directions. The data for powder sample are superimposed.}\label{Fig.S4}
\end{figure}

Refinement residuals $R_I$ and $R_p$ use standard definitions of \texttt{JANA2006}. The intensity-based $R$-value is determined via integrated reflection intensities $I_i$,
\begin{equation}
 R_I=\frac{\sum_i\left|I_i^{\rm obs}-I_i^{\rm calc}\right|}{\sum_i I_i^{\rm obs}}.
\end{equation}
In contrast, the profile-based $R_p$ gauges the agreement at each point of the powder profile,
\begin{equation}
 R_p=\frac{\sum_i\left|y_i^{\rm obs}-y_i^{\rm calc}\right|}{\sum_i y_i^{\rm obs}},
\end{equation}
where $y_i$ is the observed (calculated) intensity at a given $2\theta$ angle.

\subsection{Magnetic susceptibility}

In Fig.~\ref{Fig.S4}, we show field- and temperature-dependent magnetization measurements on an individual single crystal of K$_2$IrBr$_6$. The data are somewhat noisy because of the very small crystal mass, but clearly show the absence of magnetic anisotropy, in agreement with the $j_{\rm eff}=\frac12$ state of Ir$^{4+}$.

\subsection{Crystal-field splitting}
The effect of individual structural distortions on the crystal-field splitting $\Delta$ was studied for several model structures. They were constructed as follows:
\begin{itemize}
 \item Octahedral rotations ($\varphi$): space group $P4/mnc$, the octahedra are kept rigid and rotate in the $ab$ plane
 \item Octahedral tilts ($\psi$): space group $I2/m$, the octahedra are kept rigid and tilt in the $ab$ plane
 \item Tetragonal strain ($\eta$): space group $P4/mnc$, the Br positions are adjusted in such a way that the Ir--Br distances remain constant
 \item Tetragonal deformation ($\delta_t$): space group $P4/mnc$, the Br positions are adjusted in such a way that the average Ir--Br distance $(d_1+d_2)/2$ remains constant
 \item Orthorhombic deformation ($\delta_o$): space group $Fmmm$, the Br positions are adjusted in such a way that $(d_1+d_1')/2=d_2$ remains constant
\end{itemize}
In all cases except $\delta_o$, the tetragonal/monoclinic unit cell of $\beta$-/$\gamma$-K$_2$IrBr$_6$ was used. In the case of $\delta_o$, the unit cell of $\alpha$-K$_2$IrBr$_6$ was used but with the lower symmetry to allow for the deformation. Scalar-relativistic calculations were performed, and orbital energies of the $t_{2g}$ manifold were obtained by Wannier projections.

\subsection{Exchange couplings}
Exchange couplings were estimated for the 20\,K structure of $\gamma$-K$_2$IrBr$_6$ using superexchange theory of Refs.~\cite{rau2014} and~\cite{winter2016} with the effective Coulomb repulsion $U=1.8$\,eV, Hund's coupling $J=0.3$\,eV, and spin-orbit coupling constant $\lambda=0.4$\,eV. Individual exchange tensors are given below. All components are in units of Kelvin. For the notation of bonds, see main text. 

\begin{equation}
\mathbb J_1=\left(
\begin{array}{r @{\hspace{0.3cm}}r @{\hspace{0.3cm}}r}
 20.1 & 1.5 & -0.3 \\
 1.5 & 20.0 & 0.5 \\
 -0.3 & 0.5 & 28.9
\end{array}
\right)
\notag
\end{equation}

\begin{equation}
\mathbb J_2=\left(
\begin{array}{r @{\hspace{0.3cm}}r @{\hspace{0.3cm}}r}
 15.8 & -0.9 & 0.6 \\
 -0.9 & 15.7 & 0 \\
 0.6 & 0 & 23.2
\end{array}
\right)
\notag
\end{equation}

\begin{equation}
\mathbb J_3=\left(
\begin{array}{r @{\hspace{0.3cm}}r @{\hspace{0.3cm}}r}
 20.0 & 1.5 & -0.5 \\
 1.5 & 20.1 & 0.3 \\
 -0.5 & 0.3 & 28.9
\end{array}
\right)
\notag
\end{equation}

\begin{equation}
\mathbb J_4=\left(
\begin{array}{r @{\hspace{0.3cm}}r @{\hspace{0.3cm}}r}
 15.7 & -0.9 & 0 \\
 -0.9 & 15.8 & -0.6 \\
 0 & -0.6 & 23.2
\end{array}
\right)
\notag
\end{equation}

\begin{equation}
\mathbb J_5=\left(
\begin{array}{r @{\hspace{0.3cm}}r @{\hspace{0.3cm}}r}
 14.9 & 10.4 & 5.4 \\
 -10.3 & 21.3 & 4.6 \\
 -1.1 & -6.2 & 16.4
\end{array}
\right)
\notag
\end{equation}

\begin{equation}
\mathbb J_6=\left(
\begin{array}{r @{\hspace{0.3cm}}r @{\hspace{0.3cm}}r}
 14.4 & 8.9 & -2.1 \\
 -8.3 & 18.6 & -10.3 \\
 -4.0 & 9.3 & 13.6
\end{array}
\right)
\notag
\end{equation}

\begin{equation}
\mathbb J_7=\left(
\begin{array}{r @{\hspace{0.3cm}}r @{\hspace{0.3cm}}r}
 18.6 & 9.0 & -10.0 \\
 -8.3 & 14.4 & 5.3 \\
 10.9 & 0.8 & 13.6
\end{array}
\right)
\notag
\end{equation}

\begin{equation}
\mathbb J_8=\left(
\begin{array}{r @{\hspace{0.3cm}}r @{\hspace{0.3cm}}r}
 21.3 & 10.4 & 6.4 \\
 -10.3 & 14.9 & 0.8 \\
 -4.7 & -5.1 & 16.4
\end{array}
\right)
\notag
\end{equation}

Note that $\mathbb J_1-\mathbb J_4$ are symmetric, because inversion centers in the middle of the bonds render $\mathbf D=0$. In contrast, $\mathbb J_5-\mathbb J_8$ lack such inversion symmetry and entail sizable Dzyaloshinskii-Moriya interactions.

\end{document}